\documentclass[aps,pra,amsmath,amssymb,10pt,a4paper,floatfix,twocolumn,superscriptaddress,preprintnumbers,showpacs,longbibliography,nofootinbib]{revtex4-2}
\usepackage{hyperref,xcolor}
\definecolor{DESYCyan}{cmyk}{1,0,0,0} 
\definecolor{DESYOrange}{cmyk}{0,0.51,1,0} 
\hypersetup{
    colorlinks=true,       
    citecolor=DESYCyan, 
    linkcolor=DESYCyan, 
    urlcolor=DESYCyan   
}
\usepackage{subfigure}
\usepackage{graphicx}
\usepackage{dcolumn}
\usepackage{bm}
\usepackage{color}
\usepackage{physics}
\usepackage{tabularx}
\usepackage{qcircuit}
\usepackage{mathtools}
\usepackage{multirow}
\usepackage{orcidlink}
\usepackage{makecell}
\usepackage{xspace}
\usepackage{amsthm}
\usepackage{braket}

\usepackage[capitalise]{cleveref}
\crefname{section}{Sec.}{Sections}
\Crefname{section}{Section}{Sections}

\newcommand {\obc}{OBC\xspace}
\newcommand {\pbc}{PBC\xspace}
\newcommand {\binlog}{\ensuremath{\mathrm{log}_2}\xspace}

\begin{document}

\title{Preparation of initial states with open and periodic boundary conditions on quantum devices using matrix product states}

\author{Yibin Guo \orcidlink{0000-0003-0435-1476}}
\email{yibin.guo@desy.de}
\affiliation{Deutsches Elektronen-Synchrotron DESY, Platanenallee 6, 15738 Zeuthen, Germany}

\author{Manuel Schneider \orcidlink{0000-0001-9348-8700}}
\email{manuel.schneider@nycu.edu.tw}
\affiliation{Institute of Physics, National Yang Ming Chiao Tung University, 1001 University Road, 30010 Hsinchu, Taiwan}
\affiliation{Center for Theoretical and Computational Physics, 30010 Hsinchu, Taiwan}

\author{Takis Angelides \orcidlink{0000-0002-8639-8050}}
\affiliation{Deutsches Elektronen-Synchrotron DESY, Platanenallee 6, 15738 Zeuthen, Germany}
\affiliation{Institut für Physik, Humboldt-Universität zu Berlin, Newtonstr. 15, 12489 Berlin, Germany}

\author{Karl Jansen \orcidlink{0000-0002-1574-7591}}
\affiliation{Computation-Based Science and Technology Research Center, The Cyprus Institute, 20 Kavafi Street, 2121 Nicosia, Cyprus}
\affiliation{Deutsches Elektronen-Synchrotron DESY, Platanenallee 6, 15738 Zeuthen, Germany}

\author{C.-J. David Lin \orcidlink{0000-0003-3743-0840}}
\affiliation{Institute of Physics, National Yang Ming Chiao Tung University, 1001 University Road, 30010 Hsinchu, Taiwan}
\affiliation{Center for Theoretical and Computational Physics, 30010 Hsinchu, Taiwan}
\affiliation{Centre for High Energy Physics, Chung-Yuan Christian University, 200 Chung-Pei Road, Chung-Li District, 320314 Taoyuan, Taiwan}

\author{Yao Ting Su \orcidlink{}}
\affiliation{Institute of Physics, National Yang Ming Chiao Tung University, 1001 University Road, 30010 Hsinchu, Taiwan}

\begin{abstract}
We present a framework for preparing quantum states from matrix product states (MPS) with open and periodic boundary conditions on quantum devices. The MPS tensors are mapped to unitary gates, which are subsequently decomposed into native gates on quantum hardware. States with periodic boundary conditions (\pbc) can be represented efficiently as quantum circuits using ancilla qubits and post-selection after measurement. We derive an exact expression for the success rate of this probabilistic approach, which can be evaluated a priori. The applicability of the method is demonstrated in two examples. First, we prepare the ground state of the Heisenberg model with \pbc and simulate dynamics under a quenched Hamiltonian. The volume-law entanglement growth in the time evolution challenges classical algorithms but can potentially be overcome on quantum hardware. Second, we construct quantum circuits that generate excited states of the Schwinger model with high fidelities. Our approach provides a scalable method for preparing states on a quantum device, enabling efficient simulations of strongly correlated systems on near-term quantum computers.
\end{abstract}

\maketitle

\setcounter{tocdepth}{2}

\section{Introduction}
\label{sec:intro}
Programmable quantum simulators -- both analogue and digital -- have the potential to study systems with complexities that remain challenging for classical computational methods~\cite{cirac2012goals,georgescu2014quantum,daley2022practical,dalzell2023quantum}. By harnessing controllable quantum systems to emulate more complex ones, quantum simulations are applied in fields ranging from quantum many-body physics~\cite{monroe2021programmable,shao2024antiferromagnetic,mi2022time,fauseweh2024quantum,jafarizadeh2024recipe} to quantum chemistry~\cite{arguello2019analogue,reiher2017elucidating,bauer2020quantum,lee2023evaluating,santagati2024drug} and high-energy physics~\cite{banuls2020simulating,di2024quantum,bauer2023quantum,halimeh2025quantum}. Despite the noise and limitations of current quantum devices~\cite{preskill2018quantum,bharti2022noisy}, efforts are being devoted to achieving quantum advantage and substantial progress has recently been made in simulating complex physical models~\cite{daley2022practical,shao2024antiferromagnetic}. In particular, dynamical simulations of quantum systems promise to overcome the restrictions of classical algorithms on quantum devices. The resource-efficient preparation of the desired quantum states with high accuracy is essential as a starting point for such dynamical quantum simulations~\cite{daley2022practical,dalzell2023quantum,kim2023evidence,trivedi2024quantum,halimeh2025quantum}.

For many applications, the target states are ground states of a given Hamiltonian~\cite{mcardle2020quantum,georgescu2014quantum,daley2022practical,dalzell2023quantum}. However, on near-term quantum devices, existing state preparation schemes face various challenges and limitations. Adiabatic schemes require sufficiently slow evolution to satisfy adiabatic conditions. This is often not achievable, in particular close to phase transition points~\cite{mbeng2019quantum,wang2025imaginary}. Variational quantum algorithms (VQAs) offer an alternative, but their success depends heavily on the circuit ansatz and parameter optimization strategy~\cite{dalzell2023quantum,bharti2022noisy,cerezo2021variational}, which must balance expressivity, training feasibility, and scalability. An improperly designed ansatz not only risks insufficient representational power but also suffers from optimization challenges such as barren plateaus~\cite{mcclean2018barren,ragone2024lie,fontana2024characterizing}.

The preparation of excited states poses another important task~\cite{bauman2020toward,di2024quantum,bauer2023quantum,halimeh2025quantum,daley2022practical}. Several approaches have been proposed, with the majority based on VQAs~\cite{xu2023concurrent,ding2024ground,ciavarella2025string,parrish2019quantum,nakanishi2019subspace,higgott2019variational,kuroiwa2021penalty,jones2019variational,lee2018generalized,mcclean2017hybrid,ollitrault2020quantum,tilly2020computation,santagati2018witnessing,wen2024full}. For example, variational quantum deflation (VQD) sequentially determines excited states by introducing penalty terms that suppress the overlap with previously obtained lower-energy states~\cite{kuroiwa2021penalty,jones2019variational}. Subspace-based algorithms -- such as multistate contracted variational quantum eigensolver (MC-VQE)~\cite{parrish2019quantum}, subspace search VQE (SS-VQE)~\cite{nakanishi2019subspace}, concurrent variational quantum eigensolver (cVQE)~\cite{xu2023concurrent}, and others~\cite{ding2024ground,ciavarella2025string} -- aim to directly optimize the low-energy subspace and thereby target multiple low-lying excited states simultaneously. However, these approaches face additional challenges in preparing excited states despite the limitations inherent to VQAs. VQD suffers from the accumulation of errors, which hinder the accurate determination of higher excited states. Subspace optimization algorithms depend on the ability to design sufficiently expressive and hardware-efficient circuit ansätze, capable of accurately representing multiple excited states.

An alternative arises from tensor network states (TNS), which offer a faithful representation of states in Hilbert space, in particular for ground states and low-energy excited states of local gapped Hamiltonians~\cite{orus2019tensor,banuls2023tensor,chan2016matrix,pirvu2010matrix,perez2006matrix,schollwock2011density,bridgeman2017hand,paeckel2019time,hauschild2018efficient,xiang2023density}. Various well-established algorithms exist for determining TNS corresponding to such states~\cite{orus2019tensor,banuls2023tensor,chan2016matrix,pirvu2010matrix,perez2006matrix,schollwock2011density,bridgeman2017hand,paeckel2019time,hauschild2018efficient,xiang2023density,white1992density,vidal2003efficient,jiang2008accurate,phien2015infinite,corboz2010simulation,verstraete2004renormalization,li2024accurate,schneider2021simulating,liao2019differentiable}. Since the proposal to map matrix product states (MPS)~\cite{ostlund1995thermodynamic} to quantum circuits~\cite{schon2005sequential}, numerous works have investigated deriving quantum circuits directly from TNS without problem-specific circuit design.
Several studies and algorithms have been proposed to present, extend and improve the applicability of mapping TNS to quantum circuits~\cite{delgado2007sequential,schon2007sequential,banuls2008sequentially,ran2020encoding,barratt2021parallel,lin2021real,anand2023holographic,maccormack2021simulating,malz2024preparation,schuhmacher2025hybrid,anselme2024combining,nibbi2024block,termanova2024tensor,sugawara2025embedding,mansuroglu2025preparation,zhou2021automatically}, and  practical applications have been demonstrated~\cite{smith2022crossing,jumade2023data,gonzalez2024efficient,javanmard2024matrix,rudolph2023synergistic}. Recent deterministic schemes that utilize measurement and feedback, and propagate measurement defects through tensors in accordance with symmetry constraints, have shown particularly strong performance for preparing states with large system sizes from small building blocks~\cite{smith2023deterministic,smith2024constant,sahay2024finite,zhang2024characterizing,stephen2024preparing,sahay2025classifying,gunn2025phases,piroli2021quantum}.

Nevertheless, most existing TNS-to-circuit approaches have focused on general theoretical frameworks, the preparation of relatively simple states such as AKLT states, qudit-based representations, and strategies aimed at reducing quantum resources~\cite{delgado2007sequential,schon2007sequential,banuls2008sequentially,ran2020encoding,barratt2021parallel,lin2021real,anand2023holographic,maccormack2021simulating,malz2024preparation,schuhmacher2025hybrid,anselme2024combining,smith2023deterministic,smith2024constant,sahay2024finite,zhang2024characterizing,stephen2024preparing,sahay2025classifying,gunn2025phases,piroli2021quantum,nibbi2024block,termanova2024tensor}. From a practical standpoint, a systematic framework for preparing MPS on qubit-based quantum systems -- including excited states -- is still lacking. Addressing this gap, we develop a framework for generating quantum circuits directly from TNS representations, designed for stability, efficiency, and practical implementation. We use a divide-and-conquer approach, where the MPS tensors are first mapped into multi-qubit unitary gates, before further decomposing these into hardware-native gates.

Periodic boundary conditions (\pbc) are particularly important for preparing momentum eigenstates and faster convergence to the thermodynamic limit. Yet, realizing \pbc on qubit-based devices with TNS remains challenging. In this work, we present a practical and efficient algorithm to prepare MPS with \pbc on qubit-based systems with only $2\mathrm{log}(D)$ additional ancillary qubits. Compared to recent schemes~\cite{smith2023deterministic,smith2024constant,sahay2024finite,zhang2024characterizing,stephen2024preparing,sahay2025classifying,gunn2025phases,piroli2021quantum}, our method provides a complete and qubit-native implementation, with explicit gate decompositions and minimal ancilla requirements. We provide exact expressions for post-selection success rates of our probabilistic approach.

In this paper, we employ the sequential unitary scheme~\cite{schon2005sequential} and gate decomposition techniques based on automatic differentiation (autodiff) for tensor networks~\cite{baydin2018automatic,liao2019differentiable,chen2020automatic} to treat one-dimensional cases. The final circuit depth can, in principle, be further reduced by techniques from deterministic preparation schemes~\cite{smith2024constant,sahay2024finite,zhang2024characterizing,stephen2024preparing}.
We benchmark our framework in two representative applications: (i) preparation of the Heisenberg model ground state with \pbc. We demonstrate a growth of the entanglement entropy in the time evolution according to a volume-law, which limits classical simulations but can potentially be overcome by quantum simulations; (ii) preparation of excited states of the Schwinger model, enabling the direct preparation of target excitations without requiring the preparation of lower-energy eigenstates on the quantum device.

The remainder of the paper is organized as follows. In \cref{sec:framework}, we present our divide-and-conquer framework for initializing quantum circuits from MPS. \cref{sec:pbcCircuit} demonstrates the ground state preparation of the Heisenberg model with \pbc and subsequent quench dynamics. \Cref{sec:ExciStates} details excited-state preparation for the Schwinger model. Finally, \cref{sec:Conclusion} summarizes our findings and outlines possible directions for future work.

\section{Embedding matrix product states with open or periodic boundary conditions into quantum circuits}
\label{sec:framework}
\subsection{Introduction to matrix product states}
\label{subsec:MPS}
In this subsection, we briefly introduce MPS, originally proposed in Ref.~\cite{ostlund1995thermodynamic}, which provides a paradigmatic example of a TNS with one-dimensional topology. For a system with $N$ sites, the MPS takes the form
\begin{equation}
    | \psi \rangle = \sum_{\{\sigma_{i}, \kappa_{i}\}} A^{1, \sigma_{1}}_{\kappa_{0} \kappa_{1}} A^{2, \sigma_{2}}_{\kappa_{1} \kappa_{2}} \cdots A^{N, \sigma_{N}}_{\kappa_{N-1} \kappa_{0}}|\sigma_{1}\cdots \sigma_{N} \rangle,
    \label{eq:MPS}
\end{equation}
where ${A^{i,\sigma_{i}}_{\kappa_{i-1},\kappa_{i}}}$ denotes rank-3 local tensors. The indices $\sigma_{i} \in {1,\dots,d}$ represent the local physical degrees of freedom, and the basis states $|\sigma_{1}\cdots \sigma_{N} \rangle$ are product states built from local basis states $|\sigma_{i}\rangle$. The bond indices $\kappa_{i} \in {1, \dots, D_{i}}$, represent virtual degrees of freedom that connect neighboring tensors. The set of bond dimensions ${D_{i}}$ controls the expressive power of the MPS; in practice, a site-independent upper bound $D$ is often chosen. Note that we include the summation over $\kappa_{0}$ to account for MPS with \obc and \pbc.

The efficiency of MPS stems from their entanglement structure: an MPS with bond dimension $D$ can capture states whose bipartite entanglement entropy is bounded by $\log D$, in accordance with the entanglement area law for gapped one-dimensional systems~\cite{Hastings_2007,Ge_2016}. This property makes MPS especially suitable for approximating ground and low-energy states of one-dimensional local Hamiltonians, and well-established algorithms exist for their computation.

A TNS obtained analytically or numerically can be translated into a quantum circuit that prepares the corresponding state on a quantum device. Since quantum circuits are unitary, the tensors of the TNS must be represented by, or extended into, unitary matrices. This mapping can be carried out directly for isometric tensors~\cite{schon2005sequential,lin2021real}.
For a normalized MPS with \obc, the right-canonical form, where each local tensor is an isometry, can be constructed by applying sequential $LQ$ decompositions from the rightmost tensor inward. In contrast, for MPS with \pbc, no full canonical form exists. Nevertheless, a sequential decomposition still yields a set of isometric local tensors, except for the bond matrix that connects the first and last tensor. Performing an $LQ$ decomposition followed by a singular value decomposition (SVD) of this bond matrix produces a diagonal matrix $\Lambda$ of singular values, along with two unitary matrices that can be absorbed into the adjacent tensors. This yields the representation
\begin{equation}
    \psi^{\vec{\sigma}} = \sum_{\{\sigma_{i}, \kappa_{i}\}} \Lambda_{\kappa_{N},\kappa_{0}} Q^{1, \sigma_{1}}_{\kappa_{0} \kappa_{1}} \cdots Q^{i, \sigma_{i}}_{\kappa_{i-1} \kappa_{i}} \cdots  Q^{N, \sigma_{N}}_{\kappa_{N-1} \kappa_{N}},
    \label{eq:canonical_form}
\end{equation}
where the $\{Q^{i}\}$ satisfy the isometric condition
\begin{equation}
    \sum_{\sigma_{i},\kappa_{i}} Q^{i,\sigma_{i}\raisebox{0.25ex}{*}}_{\kappa^{\prime}_{i-1},\kappa_{i}} Q^{i,\sigma_{i}}_{\kappa_{i-1},\kappa_{i}} = \delta_{\kappa^{\prime}_{i-1},\kappa_{i-1}}.
\label{eq:isometry_condition}
\end{equation}
with $\delta_{\kappa^{\prime}_{i-1},\kappa_{i-1}}$ the Kronecker delta.

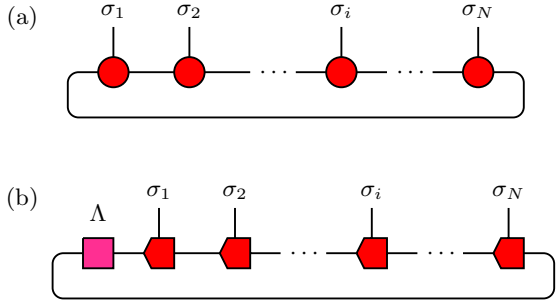
\begin{figure}[htbp!]
\centering
  \begin{tikzpicture}[every node/.style={scale=1},scale=0.4]
    \definecolor{strawberryred}{RGB}{255,47,146}{HTML}{FF2F92}
    \tikzset{every path/.style={line width=.25mm}}
    \draw (-10, 1.1) node[above] {(a)};
    \draw [rounded corners] (-7.5,0)--(-8.5,0)--(-8.5,-1.5)--(6.5,-1.5)--(6.5,0)--(5.5,0);
    \draw (-7,0.5)--(-7,1.5);
    \filldraw[fill=red, draw=black](-7,0)circle(0.5);
    \draw (-6.5,0)--(-5,0);
    \draw (-7,1.5) node[above] {$\sigma_{1}$};
    \draw (-4.5,0.5)--(-4.5,1.5);
    \filldraw[fill=red, draw=black](-4.5,0)circle(0.5);
    \draw (-4,0)--(-2.5,0);
    \draw (-4.5,1.5) node[above] {$\sigma_{2}$};
    \draw (-1.65,0) node {$\cdots$} (-1,0)--(0,0);
    \draw (0.5,1.5) node[above] {$\sigma_{i}$};
    \draw (0.5,0.5)--(0.5,1.5);
    \filldraw[fill=red, draw=black] (0.5,0) circle(0.5);
    \draw (1,0)--(2,0);
    \draw (2.85,0) node {$\cdots$} (3.5,0)--(4.5,0);
    \draw (5,0.5)--(5,1.5) node[above] {$\sigma_{N}$};
    \filldraw[fill=red, draw=black] (5,0) circle(0.5);

    \draw (-10,-4.9) node[above]{(b)};
    \filldraw[fill=strawberryred, draw=black] (-8,-5.5)--(-7,-5.5)--(-7,-6.5)--(-8,-6.5)--(-8,-5.5);
    \draw (-7.5,-5.25) node[above] {$\Lambda$};
    \filldraw[fill=red, draw=black] (-5.5,-5.5)--(-5.5,-4.5) (-6,-6)--(-5.75,-5.5)--(-5,-5.5)--(-5,-6.5)--(-5.75,-6.5)--(-6,-6) (-7,-6)--(-6,-6);
    \draw (-5.5,-4.5) node[above] {$\sigma_{1}$};
    \filldraw[fill=red, draw=black] (-3,-5.5)--(-3,-4.5) (-3.5,-6)--(-3.25,-5.5)--(-2.5,-5.5)--(-2.5,-6.5)--(-3.25,-6.5)--(-3.5,-6) (-2.5,-6)--(-1.5,-6) (-5,-6)--(-3.5,-6);
    \draw (-3,-4.5) node[above] {$\sigma_{2}$};
    \draw (-0.65,-6) node{$\cdots$} (0,-6)--(1,-6);
    \filldraw[fill=red, draw=black] (1.5,-5.5)--(1.5,-4.5) (1,-6)--(1.25,-5.5)--(2,-5.5)--(2,-6.5)--(1.25,-6.5)--(1,-6) (2,-6)--(3,-6);
    \draw (1.5,-4.5) node[above] {$\sigma_{i}$};
    \draw (3.85,-6) node{$\cdots$} (4.5,-6)--(5.5,-6);
    \filldraw[fill=red, draw=black] (6,-5.5)--(6,-4.5) (5.5,-6)--(5.75,-5.5)--(6.5,-5.5)--(6.5,-6.5)--(5.75,-6.5)--(5.5,-6);
    \draw (6,-4.5) node[above] {$\sigma_{N}$};
    \draw [rounded corners] (-8,-6)--(-9,-6)--(-9,-7.5)--(7.5,-7.5)--(7.5,-6)--(6.5,-6);
    \end{tikzpicture}
\caption{Matrix product states with (a) a general form and (b) a special form obtained after sequential $LQ$ decompositions from site $N$ to site 1. This form is analogous to the right-canonical form, except for the first diagonal bond matrix $\Lambda$. For \obc, setting the boundary index to $1$ reduces (b) to the right-canonical form.}
\label{fig:MPS}
\end{figure}

Diagrammatic notation provides an intuitive and compact way to represent TNS. For the general MPS in~\cref{eq:MPS}, the coefficient
\begin{equation}
    \psi^{\vec{\sigma}} \equiv \sum_{\{\kappa_{i}\}} A^{1, \sigma_{1}}_{\kappa_{0} \kappa_{1}} A^{2, \sigma_{2}}_{\kappa_{1} \kappa_{2}} \cdots A^{N-1, \sigma_{N-1}}_{\kappa_{N-2} \kappa_{N-1}} A^{N, \sigma_{N}}_{\kappa_{N-1} \kappa_{0}}
\end{equation}
is represented in \cref{fig:MPS}(a): red circles with 3 legs are local rank-3 tensors, open legs correspond to the physical indices $\{\sigma_{i}\}$, and connected legs imply the summation over the corresponding internal indices $\kappa_{i}$. After the sequential $LQ$ decompositions, the tensors take the form of \cref{eq:canonical_form}, shown in \cref{fig:MPS}(b). Here, the resulting isometric rank-3 local tensors are drawn as polygons to distinguish them from general local tensors, and the diagonal bond matrix $\Lambda$ is shown as a pink square. For normalized MPS with \obc, this matrix $\Lambda$ reduces to the scalar $1$, thereby recovering the right canonical form.

\subsection{Overview of the divide-and-conquer framework}
\begin{figure*}[htbp!]
    \centering
    \includegraphics[width=0.75\linewidth]{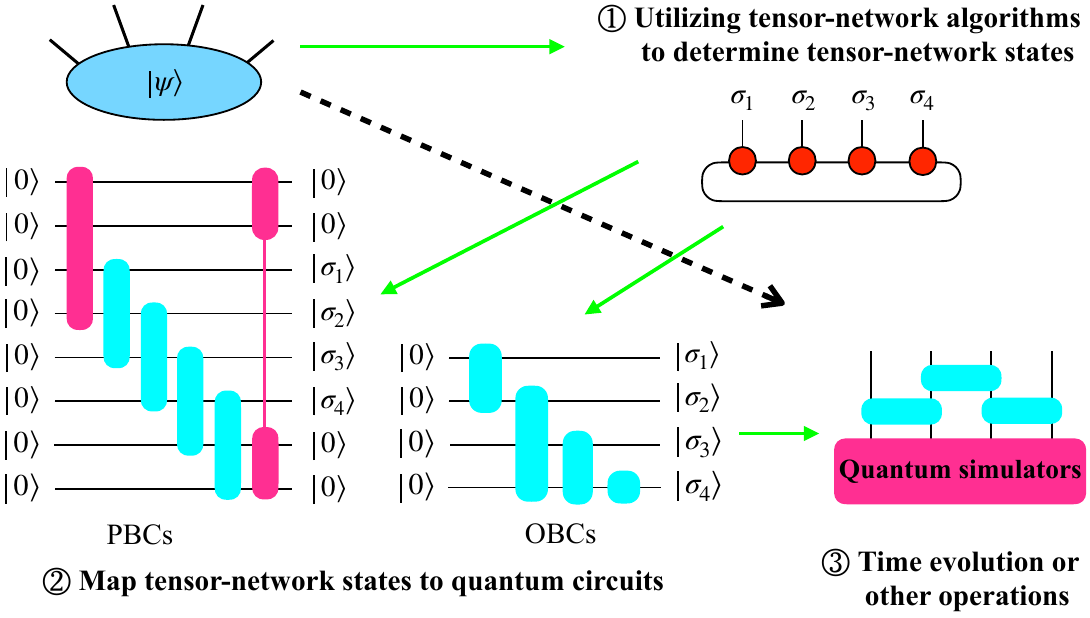}
    \caption{ Divide-and-conquer framework for the initialization of quantum circuits using matrix product states. Taking a four-qubit system as an example, the workflow involves three main steps: determining tensor-network states, mapping local tensors to quantum gates, and operating on the prepared states.}
    \label{fig:workflow}
\end{figure*}

We summarize the framework for representing an MPS by a quantum circuit in this subsection, with detailed explanations of each procedure provided in the subsequent subsection. The proposed divide-and-conquer framework aims to construct the initial state of a quantum circuit using an MPS with arbitrary boundary conditions (e.g. \obc or \pbc), as given in \cref{eq:canonical_form} or \cref{fig:MPS}(b). The procedure is illustrated for a four-site MPS in \cref{fig:workflow}. This framework consists of three main steps: the determination of the MPS representation of the desired state with the desired boundary conditions, mapping the MPS to a quantum circuit, and performing potential operations on the prepared quantum state.

The preparation of low-energy states in the context of MPS with open or periodic boundary conditions is well established~\cite{white1992density,schollwock2011density,verstraete2004density,pippan2010efficient,liao2019differentiable,vidal2004efficient,xiang2023density,li2024accurate}. This work focuses on the second step and uses the sequential unitary scheme~\cite{schon2005sequential}, with a novel treatment developed to extend its applicability to MPS with \pbc while requiring minimal modifications to the established approach. The TNS is mapped to quantum circuits by embedding isometric local tensors into unitary quantum gates with the Gram-Schmidt decomposition, followed by gate decomposition with autodiff. While the isometric structure is ensured with \obc, the additional boundary matrix for \pbc is implemented by introducing ancillary qubits. Finally, once the state is prepared, the desired operations such as a time evolution can be performed on the quantum device.

\subsection{Mapping matrix product states with \obc to quantum circuits}
\label{subsec:tensors2gatesobc}
This subsection revisits and extends the mapping from a TNS with canonical form to a quantum circuit. We present the detailed mapping procedure for MPS with \obc and further discuss the connection between TNS with canonical form and quantum circuits. We focus on a physical dimension $d = 2$, which corresponds to qubit systems.

\subsubsection{Revisiting the cases with bond dimension as power of $2$}
We begin with cases when the bond dimension is a power of $2$, enabling a direct connection to qubit gates. Once the MPS with canonical form is obtained, the local tensors become unitary or isometric after being reshaped into matrices. As explained in \cref{fig:tensor2gate}(a), unitary matrices can be directly reshaped into the form of qubit gates, while isometric matrices are equivalent to unitary tensors with a projection, a process referred to as embedding in the subsequent discussion. 

To elucidate the embedding process, we consider an isometric tensor $\mathbf{Q}^{i}$ at site $i$ and extend it to a unitary matrix $\mathbf{U}^{i}$. Without loss of generality, we assume $\mathbf{Q}^{i}$ is an isometric tensor with dimensions $D \times 2 \times D$, and further assume $D$ is a power of $2$. By combining indices $\sigma_{i}$ and $\kappa_{i}$ into a single row index, $\mathbf{Q}^{i}$ forms an isometric matrix with dimensions $2D \times D$, satisfying \cref{eq:isometry_condition}. The reshaped matrix $\mathbf{Q}^{i}$ consists of a set of orthogonal vectors, expressed as
\begin{equation}
    \mathbf{Q}^{i} = \big(q^{i}_{1}, q^{i}_{2}, \cdots, q^{i}_{D}\big)
\end{equation}
where the orthogonality condition is
\begin{equation}
    q^{i,\dagger}_{m} q^{i}_{n} = \delta_{m,n},
\end{equation}
We introduce a square matrix $\tilde{\mathbf{U}}^{i}$ with dimensions $2D \times 2D$, defined as
\begin{equation}
    \tilde{\mathbf{U}}^{i} = \big(q^{i}_{1}, q^{i}_{2}, \cdots, q^{i}_{D}, \tilde{q}^{i}_{D+1}, \cdots, \tilde{q}^{i}_{2D}\big),
\end{equation}
where $\tilde{q}^{i}$ represents vectors initialized with random numbers. The Gram-Schmidt process~\cite{leon2013gram} is then applied to $\tilde{\mathbf{U}}^{i}$, leading to orthogonal columns in the resulting unitary matrix $\mathbf{U}^{i}$. This procedure leaves the columns $q^{i}_{1}, q^{i}_{2}, \cdots, q^{i}_{D}$ unchanged and thus embeds $\mathbf{Q}^{i}$, but alters $\tilde{q}^{i}_{D+1}, \cdots, \tilde{q}^{i}_{2D}$ to form orthonormal vectors. The unitary matrix $\mathbf{U}^{i}$ can be further reshaped into an $n$-qubit gate, where $n=\binlog(D) + 1$. To recover the isometric matrix $\mathbf{Q}^{i}$, the multi-qubit gate can be applied to the $|0\rangle$ state on the last site as shown in the right panel of \cref{fig:tensor2gate}(a). This application of $|0\rangle$ is equivalent to selecting the desired vectors $(q^{i}_{1}, \cdots, q^{i}_{D})$.

\subsubsection{Decomposing multi-qubit gates into hardware-native operations}
After embedding each local isometric tensor to multi-qubit gates, the next step is to decompose these multi-qubit gates into a sequence of native gates on a quantum device. We focus on the typical case where these are single-qubit and CNOT gates. This can be done exactly using methods such as the Cosine-Sine decomposition~\cite{shende2005synthesis} or the KAK decomposition~\cite{tucci2005introduction}. However, these methods exhibit exponential scaling in both gate count and circuit depth. Rather than performing a direct decomposition, we adopt an alternative strategy by approximating the multi-qubit gates with a gate ansatz composed of universal two-qubit gates. This choice is not unique, and several gate decompositions and optimization strategies exist~\cite{rakyta2022approaching}.

\begin{figure}
    \centering
    \includegraphics[width=0.75\linewidth]{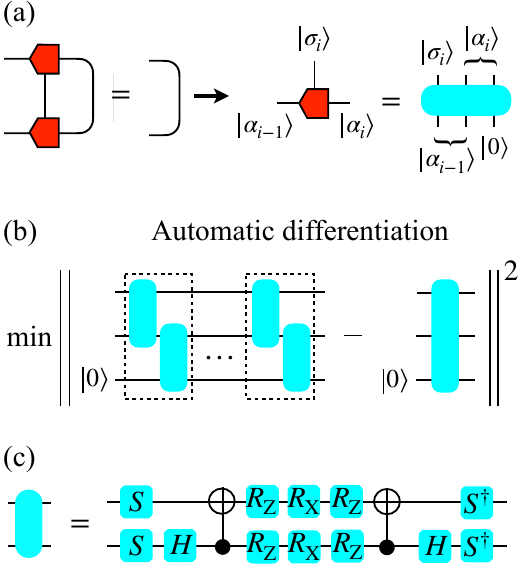}
    \caption{Schematic divide-and-conquer diagram mapping local tensors to quantum gates, using a $D=4$ MPS as an example. (a) Local tensors after sequential $LQ$ decompositions are unitary or isometric matrices after reshaping. The unitary matrices can be transformed into multi-qubit gates by reshaping them into tensors, where each index corresponds to a qubit index. For isometric matrices, a similar reshaping is applied. However, the tensors have to be extended to obtain unitary gates, and the target state is recovered by applying the initial state $\Ket{0}$ on the corresponding index.
    (b) Mapped universal multi-qubit gates are decomposed into universal $\mathrm{SO}(4)$ gates by minimizing the squared Frobenius distance between the decomposed and target gates. (c) Each universal $\mathrm{SO}(4)$ gate is further decomposed into two CNOT gates and a set of single-qubit gates.}
    \label{fig:tensor2gate}
\end{figure}

In this work, we approximate the gates by a sequence of universal $\mathrm{SO}(4)$ gates, a viable and hardware-efficient choice for real-valued MPS matrices. The determinant of each quantum gate obtained from the MPS is chosen to be +1 by a gauge transformation as explained in \cref{appendix:fix_gauge}. Then, these gates are approximated by a sequence of universal $\mathrm{SO}(4)$ gates. These are found by optimizing the Hilbert-Schmidt distance between the decomposed gates and the target gates. As shown in \cref{fig:tensor2gate}(b), we adopt a ladder-type structure with multiple layers of universal $\mathrm{SO}(4)$ gates as the gate ansatz. More general gate layouts and choices can also be explored without loss of generality. The number of layers serves as a tunable parameter, allowing for a controllable trade-off between gate depth and decomposition accuracy. We use automatic differentiation for the minimization task.

In practice, it is convenient to construct a universal $\mathrm{SO}(4)$ gate by generating a rotation matrix $G$ through the exponential map from a real skew-symmetric matrix $A$,
\begin{equation}
    G = \exp(A).
    \label{eq:exponential_map}
\end{equation}
The exponential map provides a flexible parameterization for variational optimization, enabling the convenient exploration of the entire space of two-qubit orthogonal operations. The optimized $\mathrm{SO}(4)$ gates can be further decomposed into single qubit gates and CNOT gates, as shown in \cref{fig:tensor2gate}(c). While the $\mathrm{SO}(4)$ gate ansatz is adopted to demonstrate feasibility due to its sufficient expressive ability~\cite{guo2024concurrent}, other circuit ansätze can be also explored.

\subsubsection{Extension to general bond dimensions}
We propose a way to extend the conversion of an MPS to a quantum circuit for the general case when the bond dimension is not a power of $2$. First, the MPS is compressed to a smaller bond dimension that is a power of two. For \obc, this compression can be performed directly with the techniques based on the canonical form~\cite{schollwock2011density}.  The corresponding quantum circuit for this MPS with reduced bond dimension can be obtained as discussed before. To account for the information loss in the compression process, additional gate layers are introduced and are optimized to maximize the fidelity with the initial MPS. The inverse of these gates effectively reduces the entanglement if applied to the initial MPS, and we therefore refer to these gates as disentanglers.

\begin{figure}
    \centering
    \includegraphics[width=0.8\linewidth]{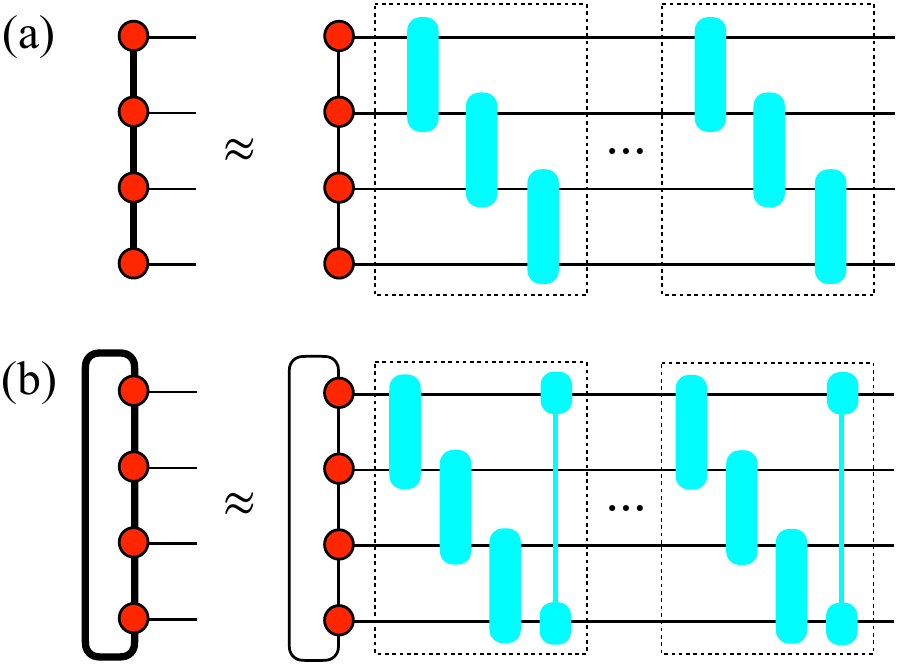}
    \caption{ The four-site examples for disentangling MPS with \obc (a) and \pbc (b) into MPS with bond dimension constrained to powers of $2$, where these MPS with smaller bond dimensions are directly obtained through compression. The disentangling process is implemented using sequential universal $\mathrm{SO}(4)$ gates arranged in a ladder structure, which can be alternatively rearranged into another structure. The universal $\mathrm{SO}(4)$ gates are optimized by minimizing infidelity to achieve effective disentanglement. }
    \label{fig:DisentangledMPS}
\end{figure}

The optimization of the additional disentangling gates is shown in \cref{fig:DisentangledMPS}(a). We minimize the infidelity between the original MPS and the compressed MPS with a sequence of universal $\mathrm{SO}(4)$ gates arranged in a ladder structure. Other configurations, such as a brick-wall structure, could be applied as well. Once the structure is fixed, we optimize the disentangling gates with different gate layers by tuning the skew-symmetric matrix A according to the exponential map in \cref{eq:exponential_map}, using autodiff. This procedure enables the mapping of an MPS with a general bond dimension to a quantum circuit.

The above processes of state compression, isometry tensor mapping, and gate decomposition provide a systematic approach for mapping MPS with \obc to quantum circuits of single-qubit and CNOT gates. Since canonical forms are available for loop-free TNS, this method can, in principle, be extended to more general cases, such as tree tensor-network states (TTN) and multi-scale entanglement renormalization ansatz (MERA).

\subsection{Mapping matrix product states with \pbc to quantum circuits}
In this subsection, we introduce the mapping strategy to MPS with \pbc. For clarity, we focus on the case where the bond dimension $D$ is a power of two. For other values of $D$, a analogous strategy to that used in the \obc case can be applied. The required compression step for \pbc is detailed in \cref{appendix:compression}. As shown in \cref{fig:DisentangledMPS}(b), the disentangling gates are optimized by minimizing the infidelity between the original and the compressed MPS with disentangled gates applied.

The embedding of an MPS with \pbc is illustrated in \cref{fig:workflow}. We assume an MPS consisting of right isometries, which can be constructed as discussed in \cref{subsec:MPS} and shown in \cref{fig:MPS}(b). The isometric tensors can be embedded in quantum gates as explained before for \obc.  However, the bond matrix $\Lambda$ is not unitary. Interpreted as an operator acting on a quantum state, each entry of the diagonal matrix multiplies a coefficient of the state by a factor, and therefore generally changes the norm of the state. The probability to find the desired state is therefore not one, as it would be after applying a unitary gate. We can implement this behavior -- together with the trace over the boundaries -- on a quantum device by applying unitary gates that couple the system to ancilla qubits, and a final measurement process. The measurement allows to incorporate the non-unitary character of the operation. The changed norm of the state manifests itself in a post-selection procedure: only if the ancilla qubits are in the $\Ket{0}$ state after the measurement, the state was prepared correctly\footnote{The choice of the desired outcome is not unique, as well as the initialization of the ancilla qubits. We choose $\Ket{0}$ for both the initialization and the desired post-selection outcome.}. Otherwise, the device needs to be reset and initialized again. The success rate of this probabilistic approach is therefore crucial for the efficiency. We explicitly show the construction of a circuit that implements the diagonal boundary matrix in a hardware-efficient way, and we derive an exact expression for the post-selection success rate. Throughout this subsection, we use the case $D = 4$ as an example to aid understanding.

\subsubsection{Encoding a diagonal bond matrix $\Lambda$ into unitary gates}
The non-negative real-valued diagonal bond matrix
\begin{equation}
    \Lambda = \mathrm{diag}(s_{1},s_{2},\cdots,s_{D})
\end{equation}
is split into two parts, each of which is implemented by unitary gates. We denote these two components as $\Lambda^{1-\alpha}$ and $\Lambda^{\alpha}$, where
\begin{equation}
    \Lambda^{x} = \mathrm{diag}(s^{x}_{1},s^{x}_{2},\cdots,s^{x}_{D})
\end{equation}
with $x = \alpha$ or $x = 1-\alpha$. 

To embed $\Lambda^{x}$ into a qubit gate, we first reshape the matrix into a normalized vector $v^{x}$, and then construct a unitary gate by extending this vector to an orthonormal basis as shown below in \cref{eq:vx}. The reshaping and normalization process yields the following element-wise correspondence between the diagonal matrix $\Lambda^{x}$ and the reshaped normalized vector $v^{x}$:
\begin{equation}
     v^{x}_{D(m-1)+n,1} = \frac{\big(\Lambda^{x}\big)_{m,n}}{C_{x}}.
     \label{eq:vx}
\end{equation}
Here, the indices $m$ and $n$ run from $1$ to $D$. The factor $C_{x}$ is given by
\begin{equation}
    C_{x} = \sqrt{\sum_{i=1}^{D} s_{i}^{2x} }
\end{equation}
and ensures that the vector $v^{x}$ is normalized. Using the Gram-Schmidt process, the unitary gate is constructed as
\begin{equation}
\label{eq:Vx}
    V_{x} = \big( v^{x}, v^{x,\perp}_{2},..., v^{x,\perp}_{D^{2}} \big),
\end{equation}
where $v^x$ from \cref{eq:vx} is placed in the first column, and $\{ v^{x, \perp}_{j} \}$ are orthonormal vectors that complete the basis. 

\begin{figure}[htbp!]
    \centering
    \includegraphics[width=0.75\linewidth]{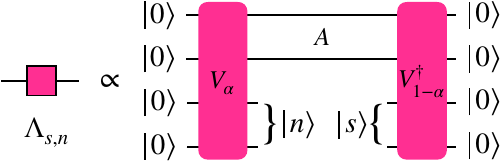}
    \caption{Encoding of the real-valued diagonal bond matrix $\Lambda$ for MPS with \pbc. The matrix $\Lambda$ is rescaled by the $C_{\alpha}C_{1-\alpha}$ and then embedded into two gates, $V_{\alpha}$ and $V_{1-\alpha}^{\dagger}$, with a final post-selection step. These two gates $V_\alpha$ and $V^\dagger_{1-\alpha}$ correspond to the first and last four-qubit gates (in pink) in the second step of the \pbc case in \cref{fig:workflow}.}
    \label{fig:bondmatrix}
\end{figure}

We illustrate the circuit encoding process using the example of $D=4$ in \cref{fig:bondmatrix}. These two gates $V_{\alpha}$ and $V_{1-\alpha}^{\dagger}$ correspond to the first and last four-qubit gates (in pink) in the second step shown in \cref{fig:workflow}. There, the right gate is acting on qubits at separate ends of the circuit and is therefore split into two parts. The open indices between the two gates in \cref{fig:bondmatrix} connect to the left-most and right-most three-qubits gates (cyan-colored) in \cref{fig:workflow}, respectively. The embedding process described above requires $2 \binlog(D) = 4$ ancillary qubits, where the first (last) two qubits in \cref{fig:workflow} correspond to the indices $m$ ($n$) of $\Lambda^{x}$. We denote the first two qubits as subsystem $A$. \Cref{fig:bondmatrix} represents the following relation
\begin{equation}
    \Lambda_{s,n} = C_{\alpha} C_{1-\alpha} \langle 0000| V^{\dagger}_{1-\alpha,s} V_{\alpha,n} |0000\rangle,
    \label{eq:bond2matrix_original}
\end{equation}
where the subscripts $s$ and $n$ are retained to indicate that these indices are not summed over during the construction. They will later be contracted with the three-qubits gates (cyan-colored) corresponding to local tensors in \cref{fig:workflow}. Identically, \cref{eq:bond2matrix_original} can be reformulated as follows. By tracing out the degrees of freedom associated with subsystem $A$, the original bond matrix $\Lambda$ can be recovered as
\begin{equation}
    \Lambda = C_{\alpha} C_{1-\alpha} \mathrm{Tr}_{A} \Big[ V_{\alpha} \mathcal{P}^{\vec{0}} V_{1-\alpha}^{\dagger} \Big],
\label{eq:bond2gates}
\end{equation} 
where the projector $\mathcal{P}^{\vec{0}} = | 0000 \rangle \langle 0000 |$ is applied. When reshaped to a matrix, this projector selects the first column of $V_{\alpha}$ and the first row of $V_{1-\alpha}^{\dagger}$, corresponding to vectors $v^{\alpha}$ and $v^{1-\alpha}$ respectively. The partial trace over $A$ is equal to multiplying $\Lambda^{\alpha}$ and $\Lambda^{1-\alpha}$, thereby reconstructing the original bond matrix $\Lambda$. As shown in \cref{fig:bondmatrix}, the implementation of the projector $\mathcal{P}^{\vec{0}}$ amounts to applying the gate corresponding to $V_{\alpha}$ on the initial state $|0000\rangle$ on the left, followed by post-selecting the outcome $|0000\rangle$ after applying the gate corresponding to $V_{1-\alpha}$ on the right. The last two qubits are left uncontracted to realize the partial trace. The indices $m$ and $n$ indicate the indices between the circuit output on the physical qubits and the entries of $\Lambda$.

\subsubsection{Generating the quantum circuit for \pbc}
The complete quantum circuit for \pbc is obtained by first bringing the tensors to right-canonical form. These tensors can then be constructed on the quantum device with the same techniques as for \obc from \cref{subsec:tensors2gatesobc}. The remaining bond matrix $\Lambda$ is finally implemented via post-selection as previously discussed. A scheme of the complete quantum circuit is shown in \cref{fig:workflow}.

We show that this construction creates the original state represented by an MPS with \pbc, up to the normalization factor $C_{x}$. We use the previously introduced example with bond dimension $D=4$. The norm of the state $|\psi \rangle$ is
\begin{equation}
    \mathcal{N}_{|\psi\rangle} = \langle \psi | \psi \rangle.
    \label{eq:MPSnorm}
\end{equation}
On the other hand, the quantum state
\begin{equation}
    |\psi\rangle_{\mathrm{circ}} = V^{\dagger}_{1-\alpha} U^{4} U^{3} U^{2} U^{1} V_{\alpha}|0\rangle^{\otimes 8},
\end{equation}
is normalized before post-selection, which implies
\begin{equation} 
\langle \psi_{\mathrm{circ}} | \psi_{\mathrm{circ}} \rangle = 1. 
\end{equation}
Inserting \cref{eq:bond2matrix_original} for the boundary gates with post selection, and using the condition $\mathbf{Q}^{i} = \mathbf{U}^{i}|0\rangle$ from \cref{subsec:tensors2gatesobc}, we obtain
\begin{equation}
    \tilde{\psi}^{\vec{\sigma}} = C^{-1}_{\alpha} C^{-1}_{1-\alpha} \sum_{\{\kappa_{i}\}} \mathbf{Q}^{1, \sigma_{1}}_{\kappa_{0} \kappa_{1}} \mathbf{Q}^{2, \sigma_{2}}_{\kappa_{1} \kappa_{2}} \mathbf{Q}^{3, \sigma_{3}}_{\kappa_{2} \kappa_{3}} \mathbf{Q}^{4, \sigma_{4}}_{\kappa_{3} \kappa_{4}} \Lambda_{\kappa_{4}, \kappa_{0}}
\end{equation}
with the coefficient defined as
\begin{equation}
\tilde{\psi}^{\vec{\sigma}} \equiv \langle 00 \sigma_{1} \sigma_{2} \sigma_{3} \sigma_{4} 00 | {\psi}_\mathrm{circ} \rangle.
\end{equation}
This coefficient equation implies that the resulting state after post-selection on the corresponding qubits yields the target MPS $|\psi\rangle$, up to a normalization factor $C_{\alpha}C_{1-\alpha}$ resulting from the rescaling of the bond matrix $\Lambda$. Comparing to \cref{eq:canonical_form} yields
\begin{equation}
    |\psi\rangle = C_{\alpha} C_{1-\alpha} \langle 0000 | \psi_{\mathrm{circ}} \rangle.
    \label{eq:MPS_circuit_states}
\end{equation}

The above derivation generalizes to arbitrary bond dimensions that are a power of $2$, and \cref{eq:MPS_circuit_states} holds in this general case.

\subsubsection{Success rate of post-selection for MPS with \pbc}
The success probability of post-selecting $\Ket{0_\text{anc}}$, where all ancillas are in the zero state, is
\begin{equation}
    P_{\mathrm{success}} = || \Braket{0_\text{anc} | \psi_{\mathrm{circ}} } ||^{2}.
    \label{eq:success_rate}
\end{equation}
Using \cref{eq:MPS_circuit_states,eq:MPSnorm}, it is given by
\begin{equation}
    P_{\mathrm{success}} = \frac{\mathcal{N}_{|\psi\rangle}}{C^{2}_{\alpha} C^{2}_{1-\alpha}}.
    \label{eq:success_rate_final}
\end{equation}

Since $\mathcal{N}{|\psi\rangle}$ is fixed for a given MPS, the only way to increase the success probability is to minimize the normalization factor $C{\alpha} C_{1-\alpha}$\footnote{A rescaling of the whole MPS does not affect the success rate because $C^{2}_{\alpha} C^{2}_{1-\alpha} \propto \mathcal{N}_{|\psi\rangle}$.}. This is achieved by a symmetric (equal) division of $\Lambda$ with $\alpha=1/2$. The optimal success probability is then given by
\begin{equation}
    P^{\mathrm{max}}_{\mathrm{success}} = \frac{\mathcal{N}_{|\psi\rangle}}{C^{4}_{1/2}}.
    \label{eq:successrate}
\end{equation}
In this case, the success rate for a normalized state with $\mathcal{N}_{|\psi\rangle} =1$ is bounded by $1 \ge P_{\mathrm{success}} \ge \frac{1}{D \sum_{i=1}^D s_i^2}$, where $P^{\mathrm{max}} = 1$ is achieved for \obc.
As shown in \cref{appendix:max_success_rate}, this result can be formally justified using the Cauchy–Schwarz inequality. We conclude that splitting $\Lambda$ into two equal parts—subsequently reshaped into vectors and embedded into the two unitary gates—achieves the optimal post-selection success rate within our construction. In the remainder of this paper, we adopt this equal-splitting scheme and, for simplicity, denote the corresponding gate $V_{1/2}$ as $V$.

Note that the success rate in \cref{eq:successrate} depends on the singular values of the bond matrix. It is larger, the faster the singular values decay. For anisotropic lattices, the cut defining the first and the last site of the periodic lattice can be chosen such that the success rate is maximized.

\subsubsection{Decomposition of boundary gates into elementary qubit gates}
We further decompose the gate $V$ in \cref{eq:Vx,fig:bondmatrix} obtained from an MPS into elementary qubit gates, providing a direct estimate of quantum resources required to construct the bond matrix. When decomposing $V$ into elementary two-qubit and single-qubit gates, two key observations enable substantial reductions in both gate count and circuit depth. First, it is sufficient to prepare only the first column of $V$ exactly, since the gate is always applied to input product states, which effectively project out all other columns. Therefore, the remaining columns of $V$ are irrelevant to the intended operation. Second, the first column of $V$ is sparse, containing only $D$ nonzero elements out of the total $D^{2}$ entries. These nonzero elements are located at positions corresponding to entangled basis states between the ancillary and physical qubits after reshaping. This fact motivates the use of entangled pairs between ancillary and physical qubits, which effectively select the positions of the nonzero elements. Based on these observations, Givens rotations are employed to match the values of the nonzero components~\cite{vartiainen2004efficient}. Furthermore, Gray code encoding is used to minimize both the circuit depth and the number of gates~\cite{vartiainen2004efficient}. Together, these techniques lead to a substantial reduction in gate complexity compared to a generic decomposition. 

We illustrate the construction steps with an example of bond dimension $D=8$, as shown in \cref{fig:Givens_rotation}, and provide more of the technical details in \cref{appendix:bond_matrix_decomposition}. For convenience, we denote Given rotations for the first column $R_{k,1}$ as $R_{k}$, and the corresponding rotation angle $\theta_{k,1}$ as $\theta_{k}$. As \cref{fig:Givens_rotation}(a) shows, the decomposition is divided into three parts, each enclosed in a dashed-line box. Starting from the product state $|0\rangle^{\otimes 6}$, seven inverse Givens rotation operators $R^{\dagger}_{k}$ are applied to the first three qubits. They encode the amplitudes of the matrix elements of $V$. Each rotation is implemented as a multi-controlled $R_{Y}$ gate, where filled (empty) control symbols indicate control on $|1\rangle$ ($|0\rangle$). The rotation angle may acquire a minus sign depending on the Gray code ordering. To reflect this, rounded (green) and rectangular (magenta) blocks are used in \cref{fig:Givens_rotation}(b) to distinguish $C^{n}R_{Y}(-\theta_{i})$ and $C^{n}R_{Y}(\theta_{i})$ gates. Next, two CNOT gates are applied to transform the basis from Gray code to standard binary code, as shown in the second dashed box in \cref{fig:Givens_rotation}(a). These two stages together construct the submatrix $U_{\mathrm{sub}}$. Finally, three additional CNOT gates are applied to ensure that the nonzero amplitudes appear only when the basis states of the first and second halves of the qubits are identical. Together, these three components yield the complete qubit-level construction of the qubit gate required in \cref{fig:bondmatrix} for a diagonal bond matrix with bond dimension $D=8$.

\begin{figure}[htbp!]
    \centering
    \includegraphics[width=0.9\columnwidth]{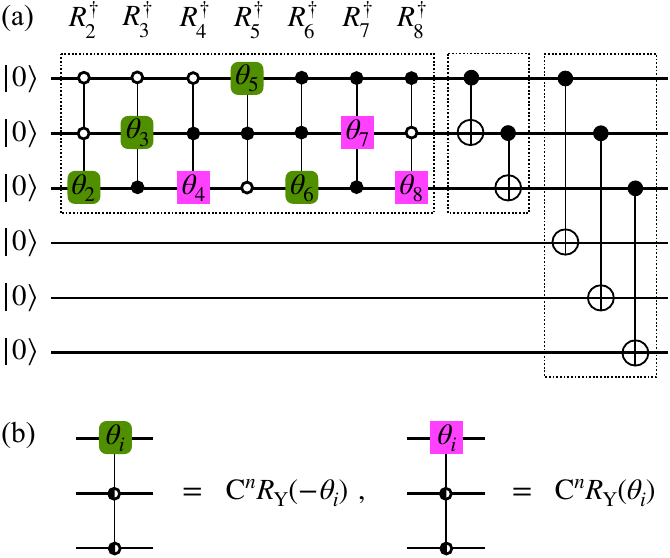}
    \caption{Decomposition of the first column of the unitary gate $V$, shown for $D=8$. (a) The first column contains nonzero elements only when the basis states of the first half of the qubits are equal to those of the second half, a structure that arises from reshaping a diagonal matrix. This decomposition is divided into three parts, highlighted in dashed boxes. First, a sequence of multi-controlled $R_{\mathrm{Y}}$ gates $\mathrm{C}^{n}R_{\mathrm{Y}}$, constructed from Givens rotations, prepares the correct amplitudes in the first-half subspace, where basis states are encoded using Gray code. Second, a basis transformation converts the Gray code to binary encoding. Finally, CNOT gates are applied to copy the state of the first half to the second half. Under Gray code encoding, some rotation angles must be flipped in sign when the local basis of an $R_{\mathrm{Y}}$ gate is ordered as $|1\rangle$, $|0\rangle$ instead of the standard order $|0\rangle$, $|1\rangle$. (b) Rounded blocks represent multi-controlled $R_{\mathrm{Y}}(-\theta_{i})$ rotations. Sharp blocks indicate multi-controlled $R_{\mathrm{Y}}(\theta_{i})$ rotations.
    }
    \label{fig:Givens_rotation}
\end{figure}

Since each $n$-qubit multi-controlled $R_{\mathrm{Y}}$ gate requires $\mathcal{O}(n)$ gates and $\mathcal{O}(n)$ circuit depth when decomposed into single qubit rotations and CNOT gates~\cite{zindorf2024efficient,vale2023decomposition}, our circuit construction provides a direct upper bound on the overall quantum resource cost. The circuit shown in \cref{fig:bondmatrix} sets an upper bound of $\mathcal{O}(D \log D)$ on both the CNOT gate depth and the total number of CNOT gates.

\subsection{Operating on the prepared states}
Once the MPS corresponding to the desired states is prepared using quantum circuits, further operations can be performed, as shown in the final step of \cref{fig:workflow}. For example, applying time-evolution gates to the circuits states enables the study of dynamics, such as quenches. This approach extends the limitation of tensor network simulations, where the typical growth of entanglement entropy and the absence of a canonical form may hinder the simulation feasibility~\cite{banuls2023tensor}. As another application, the mapped quantum circuits can serve as initial states and a starting point for a circuit ansatz in variational quantum eigensolver (VQE) algorithms~\cite{rudolph2023synergistic} or other techniques to improve the fidelity of the circuit with a desired state. The VQE in many cases suffers from a flat energy landscape~\cite{larocca2025barren}. Therefore, a good initialization close to the desired state is paramount. For example, the state shwon in \cref{fig:DisentangledMPS} could be initialized using the techniques in this paper, and then more layers can be added and optimized using VQE. This seems promising since recent results show that energy optimization over isometric TNS ansätze, such as MPS with \obc and multi-scale entanglement renormalization (MERA)~\cite{vidal2007entanglement}, are free of barren plateaus for Hamiltonians with finite-range interactions~\cite{miao2024isometric}.

\section{Applications I: quantum states preparation with \pbc for time evolution}
\label{sec:pbcCircuit}
In this section, we demonstrate how the previously introduced framework can be used to generate an initial state by a quantum circuit, and successively study quenched dynamics in a system with \pbc.  Specifically, we consider the ground state of the Heisenberg model, followed by a quench in the z-interaction. This example highlights the advantages of our state preparation framework, which integrates the efficient initialization of TNS with the capability of QC to effectively handle entanglement growth.

\subsection{Setup}
The Hamiltonian of the one-dimensional Heisenberg model with \pbc is given by
\begin{equation}
    H = J \sum_{i=1}^{N} \big( S_{i}^{x} S_{i+1}^{x} + S_{i}^{y} S_{i+1}^{y} + S_{i}^{z} S_{i+1}^{z} \big),
\end{equation}
where $J$ is the coupling constant (set to $J=1$ without loss of generality), and $S_{i}^{\alpha} = \frac{1}{2} \sigma_{i}^{\alpha}$ with $\sigma^{\alpha}$ the Pauli matrices ($\alpha = x,y,z$). Here, we identify site $N+1$ with site $1$, such that $S_{N+1}^{\alpha}\equiv S_{1}^{\alpha}$. The ground state can be represented by an MPS with \pbc as shown in \cref{fig:MPS}(b), and its determination has been addressed in previous research~\cite{schneider2021ground,schollwock2011density,xiang2023density}. In this paper, we adopted the variational optimization described in Ref.~\cite{schneider2021ground} to obtain the ground states MPS.

We then perform quench dynamics with the Hamiltonian
\begin{equation}
    \tilde{H} = \sum_{i} \big( S_{i}^{x} S_{i+1}^{x} + S_{i}^{y} S_{i+1}^{y} + \Delta S_{i}^{z} S_{i+1}^{z} \big).
\end{equation}
By using the second-order Trotter-Suzuki decomposition~\cite{Suzuki1990FractalDO,OstmeyerTrotterization}
\begin{equation}
    e^{-i\tilde{H}t} \approx (e^{-\mathrm{i}\frac{H_{o}\delta t}{2}} e^{-\mathrm{i}H_{e}\delta t} e^{-\mathrm{i}\frac{H_{o}\delta t}{2}})^{N_{t}}.
    \label{eq:Trotter}
\end{equation}
The state is evolved by sequentially applying the corresponding two-qubit gates. Here, $H_{o/e} \equiv S_{i}^{x} S_{i+1}^{x} + S_{i}^{y} S_{i+1}^{y} + \Delta S_{i}^{z} S_{i+1}^{z}$ for $i = \text{odd}/\text{even}$ is adopted to distinguish the interactions connecting odd/even bonds with their successors.

We demonstrate the state preparation introduced in \cref{sec:framework} and show numerical results as well as the success rate for the probabilistic approach. Furthermore, quench dynamics is investigated.

\subsection{Ground state preparation benchmark}
We begin by benchmarking the performance of the divide-and-conquer procedure introduced in \cref{fig:tensor2gate,fig:Givens_rotation}. For MPS with \pbc, the overall performance is determined by two key components. The first is the decomposition of the mapped multi-qubit gates corresponding to local tensors, as shown in \cref{fig:tensor2gate}. The second is the decomposition of the boundary bond matrix using multi-qubit gates combined with ancillary qubits and post-selection, as shown in \cref{fig:Givens_rotation}. For the bond matrix, we adopt the exact decomposition strategy based on Givens rotations~\cite{chai2025fermionic,chai2025towards} and Gray code introduced earlier. This approach provides favorable resource scaling of quantum resources, requiring at most $\mathcal{O}(D \log D)$ CNOT gates in both depth and gate count. In this subsection, we focus on benchmarking the decomposition of the multi-qubit gates mapped from local tensors, while the bond matrix is assumed to be prepared exactly.

\begin{figure}[!htbp]
    \centering
    \includegraphics[width=0.8\columnwidth]{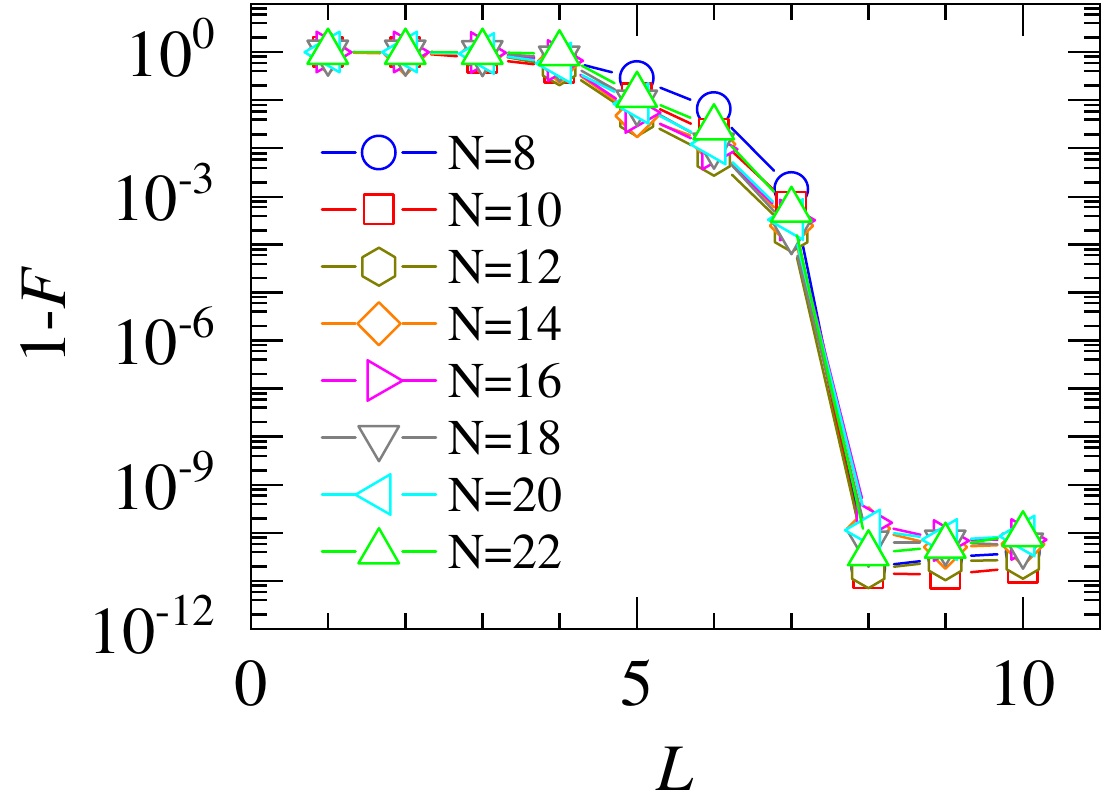}
    \caption{
    Infidelity $1-F$ between the original MPS (the ground state of the Heisenberg model with \pbc and bond dimension $D=8$) and the reconstructed state obtained by decomposing the mapped multi-qubit gates using $L$ layers of qubit gates. Results include different system sizes $N$. For all system sizes, the infidelity drops below $10^{-9}$ for $L \ge 8$.
     }
     \label{fig:HeiNscaling}
\end{figure}

To evaluate the decomposition accuracy, we examine how the infidelity scales with the number of qubit gate layers for the ground state of the Heisenberg model with bond dimension $D=8$. The results include system sizes $N = 8$ to 22 in steps of 2. Infidelity is defined as $1-F$, where $F$ is the fidelity between the original MPS and the reconstructed circuit state after decomposing the mapped multi-qubit gates. Each parameter matrix $A$ in \cref{eq:exponential_map} is initialized as a lower triangular matrix with elements chosen from a uniform distribution over $[0, 0.1)$. Optimization is carried out independently for each multi-qubit gate using the Limited-memory BFGS (L-BFGS) algorithm, with a learning rate of 0.1 and a maximum of 200 optimization steps, and a line search strategy set to the strong Wolfe condition~\cite{liu1989limited}. To mitigate the effect of local minima, we perform 10 independent optimization runs with random initial seeds and select the best result based on the lowest achieved infidelity. However, we found that the simulations are stable under different initializations and are usually not stuck in local minima. Instead of choosing the maximal fidelity, the smallest energy measured on the quantum device could also serve as a criteria if the best state shall be selected.

For all tested system sizes, the infidelity drops below $10^{-9}$ after eight layers of qubit gates are used, as shown in \cref{fig:HeiNscaling}. This indicates that at most eight qubit layers are required to accurately decomposition the MPS with $D=8$. Additionally, we tested the case of a $D=4$ and observed that the infidelity falls below $10^{-10}$ after four layers of qubit gates are used, for all system sizes chosen as in the $D=8$ case. These results confirm the effectiveness of the decomposition procedure in preparing ground state of the Heisenberg model with bond dimensions that are powers of two. We note a sudden drop to very small infidelities when the number of layers reaches $L=8$ for $D=8$. In this regime, the number of parameters in the ansatz $3 \cdot 6 L = 144$ (see \cref{fig:tensor2gate}) becomes sufficient to describe a four-qubit gate with real parameters applied to the zero state on one qubit, which has $2^{2 \cdot 4 - 1} = 128$ parameters. Therefore, the ansatz might be able to describe the multi-qubit gate exactly. This could be checked explicitly with a dimensional expressivity analysis, and redundant parameters could further be removed for a more efficient circuit~\cite{Hartung:2022cy,Funcke:2021jcs,Funcke:2021aps}.

\begin{figure}[!htbp]
    \centering
    \includegraphics[width=0.8\columnwidth]{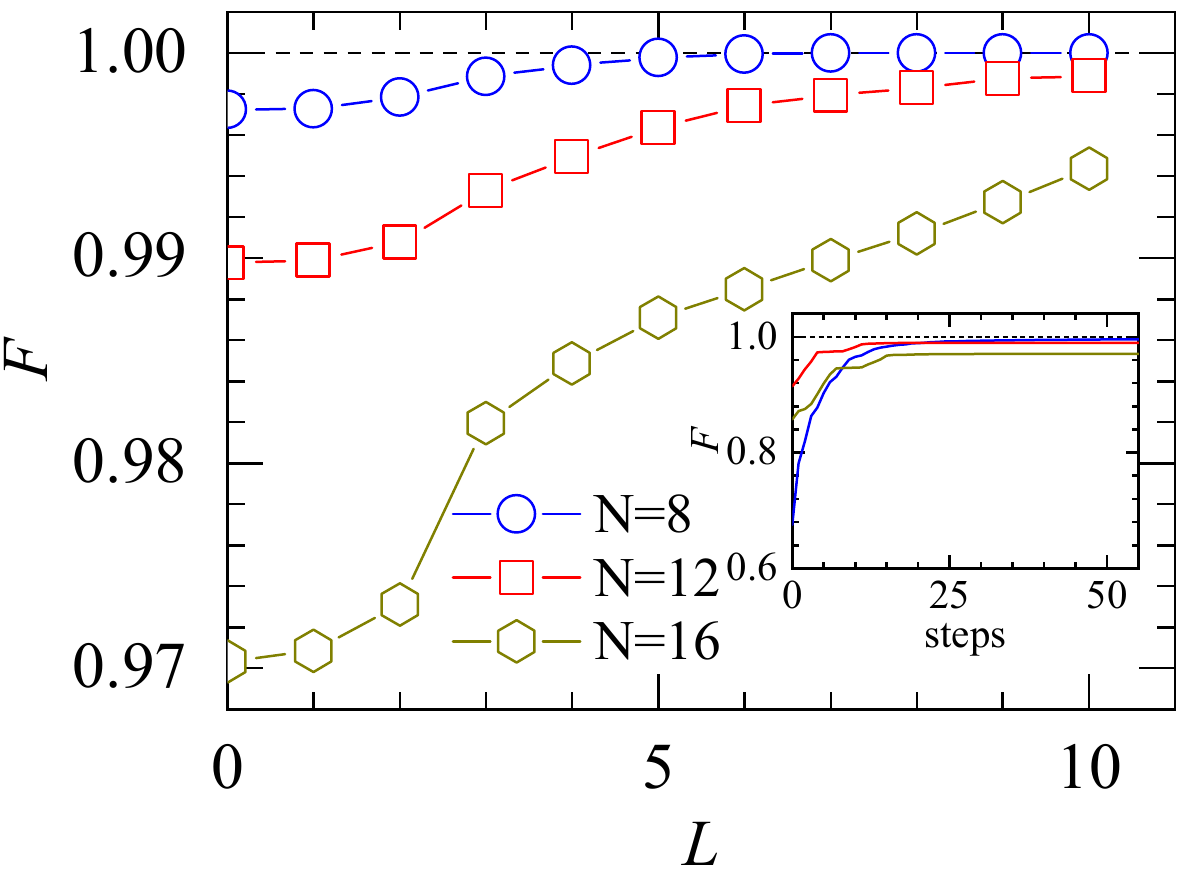}
    \caption{
    Disentangling process for MPS with $D=6$. The plot shows the fidelity $F$ between the original MPS with $D=6$, after applying $L$ layers of variationally optimized disentangling gates, and the compressed MPS with $D=4$. Results for three system sizes $N=8, 12, 16$ are given. The dashed line at $F=1$ serves as a visual guide. The inset shows the convergence of the fidelity $F$ for $L=10$ during the update process in the compression explained in \cref{appendix:compression}. Each step corresponds to a single local tensor updating step from one end of the MPS to the other.
     }
     \label{fig:HeiDisentangled}
\end{figure}

Next, we investigate the disentangling process for MPS with general bond dimensions. We consider two representative cases: $D=6$ and $D=10$. The compression strategy to a smaller bond dimension that is a power of two for \pbc is explained in \cref{appendix:compression}. For $D=10$, the compressed state with $D=8$ achieves a final fidelity exceeding 0.9998. Since this already yields a high-fidelity approximation, additional disentangling optimization provides only limited improvement. Therefore, we focus on $D=6$, where we first apply 100 full update sweeps as explained in \cref{appendix:compression}. Then, we refine the disentangling gates by minimizing the infidelity $1-F$ with 300 optimization steps, where $F$ denotes the fidelity between the original MPS with the disentangling gates applied, and the compressed MPS. The optimization procedure follows the same setup as in the ground state preparation task described previously. As before, each parameter matrix $A$ in \cref{eq:exponential_map} is initialized as a lower triangular matrix with nonzero entries, but sampled uniformly from the interval $[0, 0.01)$. Again, the best of 10 independent optimization runs is selected.

\Cref{fig:HeiDisentangled} shows the fidelity $F$ as a function of the number of disentangling layers $L$. The fidelity increases monotonically with the number of layers, indicating the effectiveness of the disentangling procedure. The inset of \cref{fig:HeiDisentangled} illustrates the convergence behavior during the first several local tensors updating steps, where each step corresponds to optimizing a single local tensor from one end of the MPS to the other. These results demonstrate that the ground state of the Heisenberg model can be effectively disentangled using the universal $SO(4)$ gate ansatz introduced in \cref{fig:DisentangledMPS}(b). Thereby, an efficient circuit-based implementation of MPS with general bond dimensions on quantum devices can be achieved.

\begin{figure}[!htbp]
    \centering
    \includegraphics[width=0.8\columnwidth]{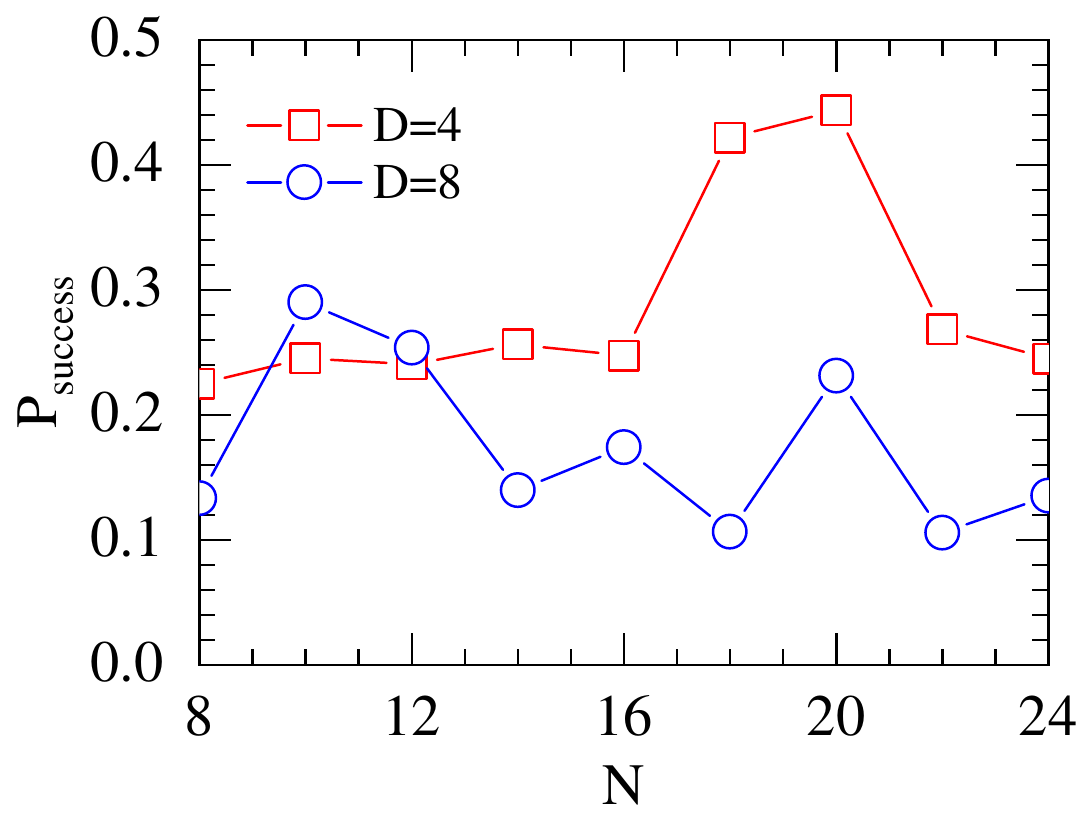}
    \caption{ Success rate $P_{\mathrm{success}}$ for preparing a normalized state with \pbc at $D=4$ and $D=8$, for system sizes $N$ ranging from $8$ to $24$ in steps of $2$. }
    \label{fig:success_rate}
\end{figure}

We further investigate how the success rate in \cref{eq:success_rate_final} changes with the system size $N$. \Cref{fig:success_rate} shows representative results for bond dimensions $D=4$ and $D=8$. The success rate $P_{\mathrm{success}}$ does not decrease monotonically with system size but instead exhibits oscillatory behavior. In all cases, $P_{\mathrm{success}}$ remains at $\mathcal{O}(10^{-1})$, indicating a sufficiently high success rate and demonstrating the effectiveness of our approach.

\subsection{Quench dynamics}
We apply the divide-and-conquer framework to simulate quench dynamics in the Heisenberg model as an example application. After preparing the ground state MPS with $D=8$, we perform a quench by changing the $J_{z}$ coupling to a new value $\Delta$, with $\Delta$ ranging from $-1.0$ to $0.8$ in steps of $0.2$. Time evolution is implemented using the second-order Trotter-Suzuki decomposition \cref{eq:Trotter} with a time step size $\delta t = 0.05$ up to a final time $t = 100$. All simulations are carried out using tensor networks with sufficiently large bond dimensions. While this is an exact simulation of an error-free quantum device, it demonstrates the entanglement growth. Thus, the simulation shows the limits where tensor networks cannot capture the exact time evolution, and quantum devices could potentially provide an advantage over classical computing resources.

\begin{figure}[!htbp]
    \centering
    \includegraphics[width=\columnwidth]{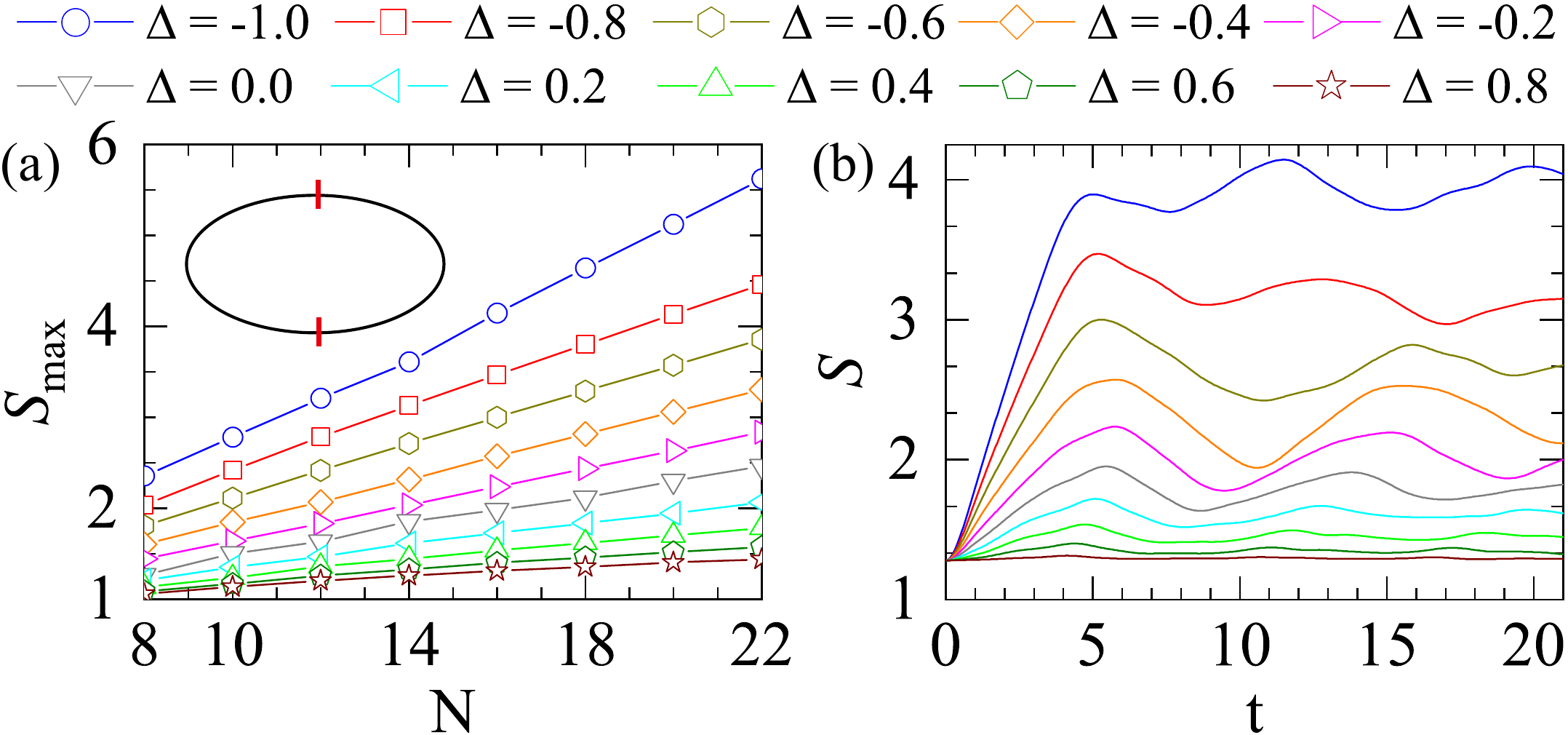}
    \caption{
     (a) Maximal entanglement entropy $S$ during time evolution as a function of the system size $N$ for different quench parameters $\Delta$, demonstrating volume law scaling. The initial states are obtained from an MPS with bond dimension $D=8$, where eight qubit layers are used to decompose the multi-qubit gates that correspond to the local MPS tensors. The inset illustrates the partitioning of the system into two equal parts in the definition of entanglement entropy. (b) Typical evolution of the entanglement entropy for a system of size $N=16$ with different quench parameters $\Delta$. 
     }
     \label{fig:XXZQuench}
\end{figure}

During the evolution, we monitor the bipartite entanglement entropy $S$, defined by partitioning the system into two equal halves shown in the inset of \cref{fig:XXZQuench}(a). This observable provides a key indicator for estimating the costs of the simulation using tensor network methods on a classical computer. \Cref{fig:XXZQuench}(a) shows the maximal entanglement entropy $S_{\mathrm{max}}$ observed during the evolution up to time $t=100$ as a function of the system size $N$ for various quench parameters $\Delta$. The entanglement entropy $S_{\mathrm{max}}$  scales linearly with $N$ for all $\Delta$, consistent with volume-law entanglement entropy growth~\cite{rigol2007relaxation,vidmar2016generalized,alba2017entanglement}. \Cref{fig:XXZQuench}(b) shows the time evolution of the entanglement entropy $S$ for $N=16$ with different quench parameters $\Delta$. We find that $S$ grows linearly and saturates at late times with oscillations.

The volume law growth in entanglement entropy that can be found in many systems~\cite{rigol2007relaxation,vidmar2016generalized,alba2017entanglement} demonstrates that tensor networks cannot follow the time evolution of quench dynamics for large system sizes and long time scales. Therefore, our initial state preparation can potentially be leveraged to overcome these limitations by simulating the dynamics on quantum hardware.

\section{Applications II: highly excited states preparation}
\label{sec:ExciStates}
We further demonstrate how excited states can be efficiently prepared on a quantum device. The Schwinger model serves as an example to show the applicability.

\subsection{Setup}
The massive Schwinger model~\cite{SchwingerOriginal,SchwingerModelMassive} describes $1+1$ dimensional quantum electrodynamics (QED) and shares many features with quantum chromodynamics (QCD) despite its relative simplicity, making it an ideal toy model for testing newly developed methods. The excited states of the Schwinger model correspond to mesons or multi-particle states, and preparing them enables the simulation of various dynamical processes using near-term quantum devices~\cite{Schwinger_massSpectrum,guo2024concurrent,partondistributionfunctionsschwinger}.

The rescaled spin Hamiltonian in dimensionless form~\cite{Banks1976,Crewther1980,Irving1983,Hamer1997,Schwinger_massSpectrum,Schwinger_QC,guo2024concurrent,partondistributionfunctionsschwinger} reads
\begin{align}
    W =\ & x\sum_{n=0}^{N-2}\left[\sigma_{n}^{+}\sigma_{n+1}^{-}+\sigma_{n}^{-}\sigma_{n+1}^{+}\right]+\frac{\mu}{2}\sum_{n=0}^{N-1}\left[1+\left(-1\right)^{n}\sigma_{n}^{z}\right] \nonumber \\
    & +\sum_{n=0}^{N-2}\left[l+\frac{1}{2}\sum_{k=0}^{n}\left(\left(-1\right)^{k}+\sigma_{k}^{z}\right)\right]^{2}.
    \label{eq:SpinHamiltonian}
\end{align}
Here, $a$ is the lattice spacing, $x=1/g^2a^2$, and $\mu=2m_{\mathrm{lat}}/g^2a$ with g the coupling constant and $m_{\mathrm{lat}}$ the fermion mass. Here, $l=\theta/2\pi$ is introduced to represent the static background electric field. We study the excited states of the system with \obc. In our simulation, we employ the sequential computation method, which determines states successively from lower to higher energy levels, as explained in Ref.~\cite{Schwinger_massSpectrum}. Our implementation uses the open-source Julia package ITensor~\cite{fishman2022itensor}. We use U(1) quantum numbers to restrict the simulations to chargeless states only. The maximal bond dimension of all MPS are set to $40$. This is sufficient to ensure that higher excited states can be found reliably in the subspace orthogonal to the previous states. The MPS for all excitations are further compressed to smaller bond dimensions using standard algorithms~\cite{xiang2023density,schollwock2011density}, and serve as the starting point for the subsequent step of mapping TNS to quantum circuits.

\subsection{Preparing excited states}
In this subsection, we present example applications of our framework in preparing excited states of the Schwinger model. We consider cases with both $l = 0$ and $l \neq 0$, and system sizes $N = 12$ and $N = 24$. The simulations are performed at fixed physical volume $\frac{N}{\sqrt{x}}=10$ and fixed lattice mass $\frac{m_{\mathrm{lat}}}{g}=0.125$,  which correspond to $x = 1.44$ and $\mu=0.3$ for $N = 12$, and $x = 5.76$ and $\mu = 0.6$ for $N = 24$. Following the same procedure used for the Heisenberg model, we begin by examining the scaling behavior of the infidelity between the target MPS and the reconstructed circuit state as a function of the number of qubit gate layers. The optimization strategy is identical to that employed for the Heisenberg model ground state, except that the number of optimization steps is set to 100. For MPS with bond dimension $D = 4$, we find that the infidelity drops below $10^{-11}$ after four layers of qubit gates are used, for all tested cases. For the $D = 8$ case, the infidelity falls below $4 \times 10^{-7}$ after eight layers, and further decreases to below $2 \times 10^{-8}$ when nine layers are used. In both cases we enter a regime where the number of parameters in the ansatz is sufficient to express the mutli-qubits gate exactly.

\begin{figure}[!htbp]
    \centering
    \includegraphics[width=0.8\columnwidth]{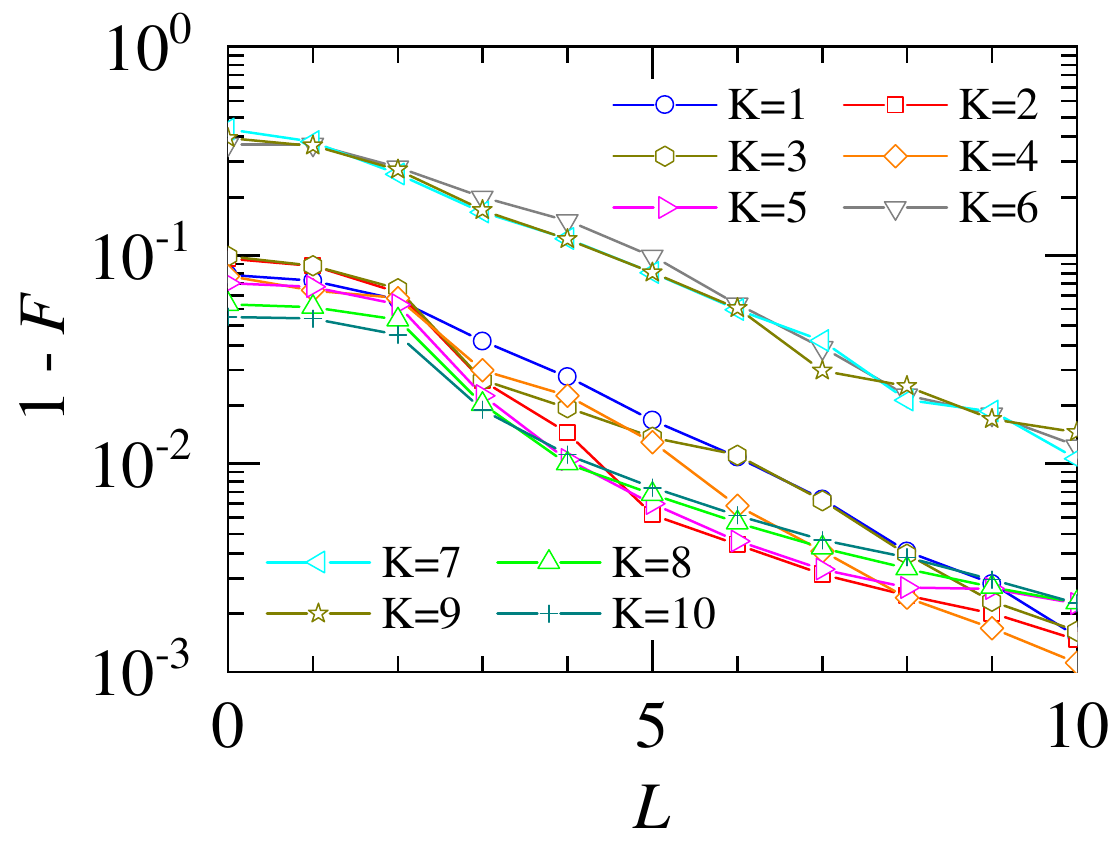}
    \caption{ Disentangling behavior of excited-state MPS for the Schwinger model with \obc. Infidelity $1-F$ as a function of the number of disentangling gate layers $L$, where $F$ is the fidelity between the compressed MPS with bond dimension $D=4$ and the original MPS with $D=6$ after applying variationally optimized disentangling gates. The results are shown for the first 10 excited states of the Schwinger model, enumerated by K, with parameters $N=16$, $x=2.56$, $\mu=0.4$, and $l=0$.}
    \label{fig:SchwingerDisentangled}
\end{figure}

The disentangling process shown in \cref{fig:DisentangledMPS}(a) is employed when the bond dimension of the target MPS is not a power of two. To evaluate its effectiveness, we compute the fidelity between the compressed MPS with disentangling gates applied and the original MPS as a function of the number of disentangling gates layers. The optimization strategy follows the same approach used for the Heisenberg model. \Cref{fig:SchwingerDisentangled} shows the infidelity $1 - F$ as a function of the number of disentangling layers $L$ for the excited-state MPS of the Schwinger model with \obc. We consider the first ten excited states at parameters $N = 16$, $x = 2.56$, $\mu = 0.4$, and $l = 0$. For $D = 10$, the fidelity after compressing to $D = 8$ remains above $0.995$, indicating a good approximation. To better illustrate the disentangling process, we consider an original MPS with bond dimension $D = 6$. As shown in \cref{fig:SchwingerDisentangled}, the 6th, 7th, and 9th excited states show the most pronounced reduction in infidelity. The infidelity decreases from values above 0.3 to below 0.02 as $L$ increases. For the remaining states, the infidelity decreases from $\mathcal{O}(10^{-1})$ to $\mathcal{O}(10^{-2})$ with increasing $L$. These results show an exponential decrease of the infidelity with the number of layers, and demonstrate the effectiveness of the disentangling procedure in excited-state MPS with \obc using the gate-based ansatz shown in \cref{fig:DisentangledMPS}(a).

\begin{figure}[!htbp]
    \centering
    \includegraphics[width=0.8\columnwidth]{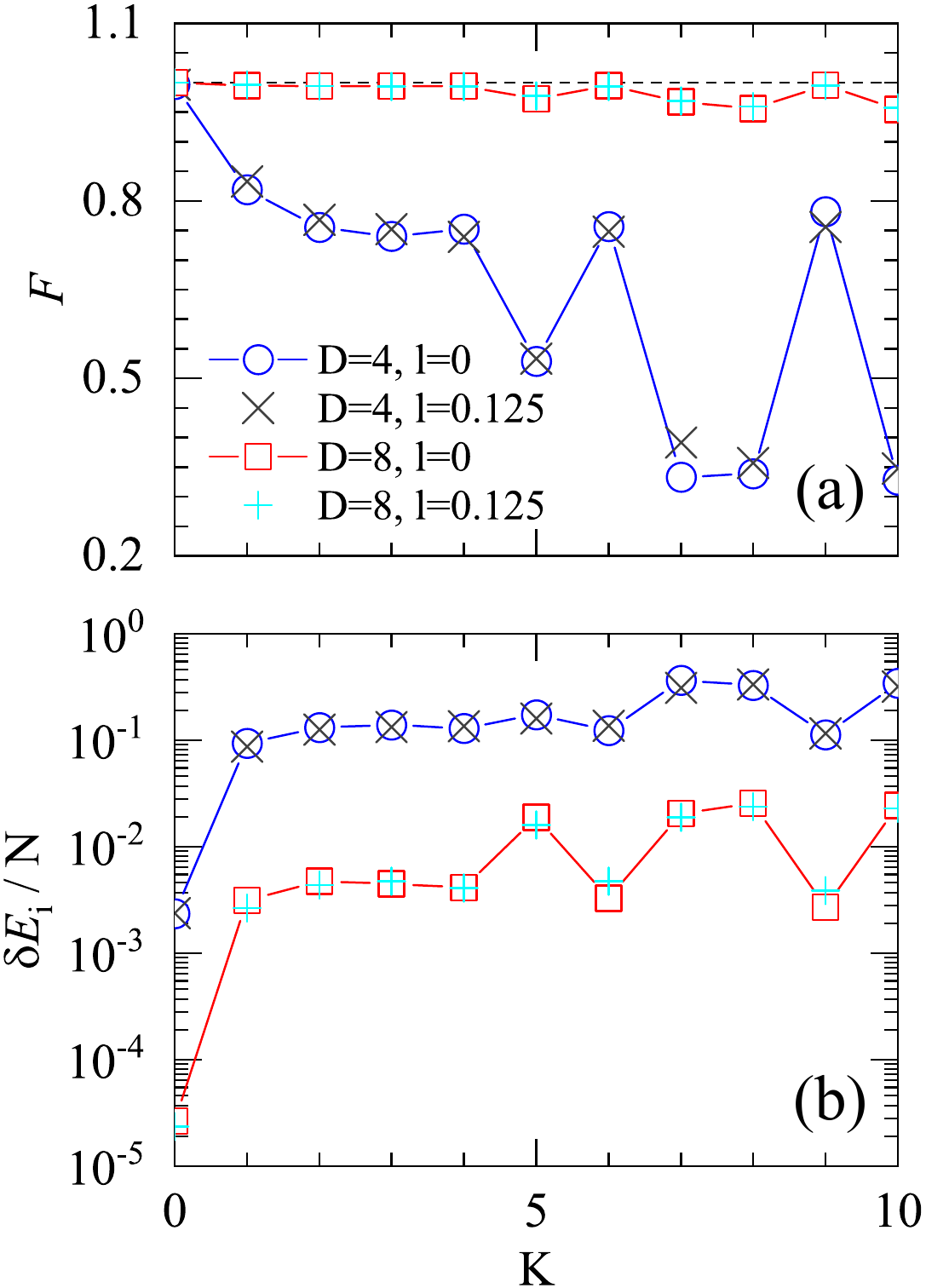}
    \caption{Comparison between the mapped and decomposed quantum circuits, and the reference states generated by MPS variational algorithms with a bond dimension $D=40$. Shown are the values for the $K$th excitation of the Schwinger model with $N=24$, $x=5.76$ and $\mu=0.6$. (a) Fidelity between the quantum circuit and the reference state. $F(D=4)$ and $F(D=8)$ correspond to circuits mapped from MPS with bond dimensions $D=4$ and $D=8$, respectively. (b) Energy difference per site $E_{\mathrm{diff}}/N$. For $D=4$, the mapped three-qubit gates are decomposed into four layers of universal $\mathrm{SO}(4)$ gates, while for $D=8$, the mapped four-qubit gates are decomposed into eight layers. With these decompositions, the squared Frobenius distance between the decomposed and target gates remains $\ge \mathcal{O}(10^{-6})$.}
    \label{fig:fidelity_Schwinger}
\end{figure}

\Cref{fig:fidelity_Schwinger} summarizes the fidelity between the quantum states from mapped and decomposed circuits, and reference states generated using MPS variational algorithms with a bond dimension $D=40$. For each excited state, we compare the fidelity $F(D=4)$ and $F(D=8)$ between the circuit-prepared states and the reference MPS, corresponding to MPS input bond dimensions $D = 4$ and $D = 8$, respectively. For the smaller bond dimension, the ground state can be approximated well while higher excited states have a lower fidelity of about $0.7$. Some higher excited states cannot be represented as well and have lower fidelities. Once an MPS with $D=8$ is chosen for the mapping to quantum circuits, the fidelity remains and close to 1 -- with $0.954450$ the worst value, a deviation of less than $5\%$. This demonstrates that the initial state construction presented in this work can generate higher excited states with high fidelities, once the bond dimension of the initial MPS is chosen sufficiently large. Furthermore, we do not observe an accumulation of errors with increasing excitation. This can be contributed to the fact that we can generate the states with high bond dimensions on classical computers before compression and mapping. Additionally, we report the per-site energy difference $E_{\mathrm{diff}}/N$ in \cref{fig:fidelity_Schwinger}(b) to quantify how well the decomposed circuits reproduce the physical energy spectrum. The energy follows a similar pattern as the fidelity: the ground state can be approximated very well even with a small bond dimension, while higher excitations have larger energy errors. Increasing the bond dimension from $D=4$ to $D=8$ decreases the error for excited states by 1 to 2 orders of magnitude. The energy can be reproduced with an error of the order of $10^{-2}$, sufficient for many applications and improvable by increasing the bond dimension further. The results demonstrate that the higher excited states can be generated directly on a quantum device with high fidelities and low errors on observables like the energy. The method avoids the accumulation of errors for higher excited states that is expected for certain VQE methods~\cite{kuroiwa2021penalty,jones2019variational}. Our approach is systematically improvable by increasing the bond dimension and therefore the circuit depth.

\section{Conclusion and Outlook}
\label{sec:Conclusion}
In this work, we systematically revisited and extended a divide-and-conquer framework for mapping MPS to quantum circuit states. We started from the previously established sequential scheme for MPS with \obc, where the canonical form plays a central role. For unitary gates corresponding to isometric local tensors in canonical form, we employed autodiff to variationally decompose the multi-qubit gates into sequences of elementary quantum gates. We established a similar formalism for MPS with \pbc, an extension which maps a TNS without canonical form to a quantum circuit. Our approach is based on the observation that, after applying a sequential $LQ$ decomposition from the right, the local tensors of MPS with \pbc retain an isometric structure, except for a boundary matrix. To represent the latter in a quantum circuit, we proposed an encoding method based on ancillary qubits and post-selection. We further analyzed the post-selection success rate and showed how it can be calculated for a given MPS from the singular value spectrum. We derived an exact formula for the success rate, which can thus be calculated from a given MPS with \pbc. We constructed an exact decomposition for the bond matrix using Gray code and Givens rotations, with a complexity of $\mathcal{O}(D \log D)$ in both the CNOT gate depth and count. To generalize the framework to arbitrary bond dimensions, we integrated an autodiff-based state disentangling algorithm. It reduces the bond dimension to the nearest power of two, allowing us to map between isometric local tensors and multi-qubit unitary gates.

To demonstrate the applicability of the framework, we presented two examples where quantum circuit states are initialized from MPS. In the first application, we prepared the ground state of the Heisenberg model with \pbc. Then, we performed quench dynamics by evolving under a time-independent Hamiltonian with varying $J_{z}$ terms, using a Trotter-Suzuki decomposition. The bipartite entanglement entropy was observed to grow over time. At late times, the maximal bipartite entanglement entropy exhibits a linear dependence on system size, indicating volume-law scaling. This behavior suggests a potential regime of quantum advantage in simulating quench dynamics in such systems. To overcome the classical limitations in entanglement entropy on a quantum device, an efficient state preparation is paramount. We demonstrated how this is possible in a hybrid approach based on MPS and showed how states with \pbc can be implemented on quantum devices.
In a second application we prepare higher excited states of the Schwinger model. High fidelities were achieved for a system of 24 sites and for the first 10 excitations. This highlights the potential for simulating excited states in quantum many-body systems in a hybrid algorithm, which does not accumulate errors for higher excitations on the quantum hardware.
Together, these applications demonstrate the generality and scalability of our framework for preparing MPS with both \obc and \pbc on quantum circuits.

Further topics in initial state preparation remain for future exploration. One promising direction is to combine the bond matrix construction method developed in this work with recently proposed measurement-and-feedback preparation techniques~\cite{malz2024preparation}. Such hybrid approaches could reduce the circuit depth from $\mathcal{O}(N)$ scaling required here to $\mathcal{O}(\mathrm{log}N)$ for MPS with $N$ sites. This reduction would make the framework appealing for near-term quantum devices and merits further investigation into its practical implementation. The resource costs can also be lowered if further structure is given, such as the block-diagonal form of MPS tensors that arises from symmetries. The bond matrix implementation presented here may serve as an initial step towards preparing higher-dimensional tensor-network states without canonical forms, such as projected entangled pair states (PEPS)~\cite{verstraete2004renormalization} and projected entangled simplex states (PESS)~\cite{xie2014tensor}, on quantum devices~\cite{banuls2008sequentially,slattery2021quantum,wei2022sequential,yu2024dual,maccormack2021simulating}.

Finally, our framework can be readily incorporated in existing and future algorithms on quantum devices to efficiently implement initial states. These can be further improved on quantum hardware to higher fidelities, or be directly utilized for applications like dynamical simulations where quantum devices can potentially overcome classical limitations.

\section*{Acknowledgments}

Y.G. is grateful to Dr.~Yahui Chai for the helpful discussions. This work is supported with funds by the European Union's Horizon Europe Framework Programme (HORIZON) under the ERA Chair scheme with grant agreement no.\ 101087126, by the Ministry of Science, Research and Culture of the State of Brandenburg within the Centre for Quantum Technology and Applications (CQTA), and by Taiwanese NSTC Grant No. 113-2119-M-007-013. T.A. is partly funded by the European Union’s Horizon 2020 Research and Innovation Programme under the Marie Sklodowska-Curie COFUND scheme with grant agreement no.\ 101034267.

\begin{center}
    \includegraphics[width = 0.125\textwidth]{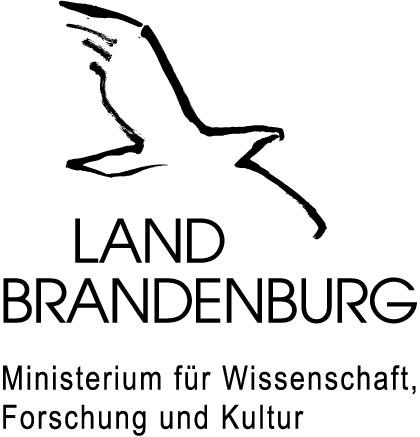}
\end{center}

\section*{Data Availability}
The data that support the findings of this study are available from the corresponding authors upon reasonable request.

\appendix

\section{Fixing the determinant of isometric MPS tensors}
\label{appendix:fix_gauge}
\begin{figure}[htbp!]
    \centering
    \includegraphics[width=0.9\linewidth]{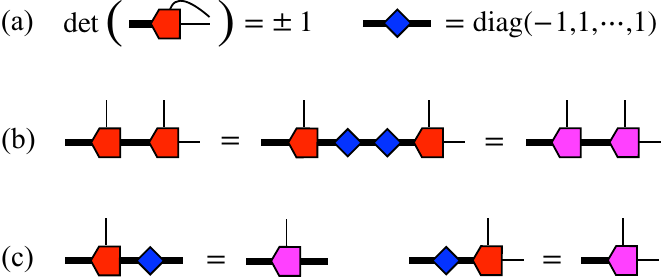}
    \caption{Ensuring determinant +1 of quantum gates. (a) After extending local MPS tensors to unitary matrices, their determinants may be either $+1$ or $-1$. If the determinant is $-1$, a gauge transformation is required to ensure compatibility with the $\mathrm{SO}(4)$ gate ansatz. For unitary MPS tensors, this is achieved by inserting a pair of diagonal matrices, each differing from the identity only in the sign of the first diagonal element, which is set to $-1$. (b-c) The inserted pair of matrices updates the neighboring local tensors. After the update, the reshaped matrix representing the right tensor has determinant $+1$. The gauge fixing proceeds from right to left until reaching a tensor whose reshaped matrix is an isometry rather than a full unitary. For the latter, the sign of the rows or columns which are added to extend the isometry to a unitary can be chosen arbitrarily without affecting the embedded MPS tensor. This ensures that the resulting gate has determinant +1.}
    \label{fig:gauge_fixing}
\end{figure}

As explained in the main text, each local MPS tensor is mapped to a sequence of universal $\mathrm{SO}(4)$ gates. Since the latter have determinant 1, this needs to be ensured for the gates to be decomposed as well. The process is illustrated in \cref{fig:gauge_fixing}. We consider real-valued MPS in this work. If the determinant of a gate is -1 instead of +1, the sign of one row or column of the corresponding gate has to be flipped. In cases where an isometric MPS tensor was embedded in a larger unitary gate, the sign flip can be applied to one of the rows or columns that were added in the extension to a unitary. The projection from the gate to the MPS tensor is not affected by such a sign flip. Close to the boundaries with \obc, the MPS tensors become unitaries, however, and no such extension exists. In this case, the sign of one row or column can be flipped together with a corresponding sign flip on one of the neighboring MPS tensors, ensuring that the physical state after contracting internal indices remains unchanged. Finally, the sign of a boundary tensor of an MPS with \obc can be flipped, corresponding to a physically irrelevant sign change of the MPS. Alternatively, one can start from both boundaries and fix the signs, until the bulk is reached where the extended rows or columns can be flipped without affecting the MPS. This way the sign of the MPS remains unchanged.

However, near the boundary of the MPS, local tensors often map to full unitary matrices. In these cases, it becomes necessary to ensure that each MPS matrix has determinant $1$. Without this condition, the gate cannot be represented using a circuit ansatz composed entirely of $\mathrm{SO}(4)$ gates. For real-valued MPS tensors as considered in this work, the determinant of the reshaped matrix is always either $+1$ or $-1$. In the latter case, the sign of the first row or column can be flipped to ensure that the determinant becomes $1$.
This can be achieved by a gauge transformation of two neighboring MPS tensors, which preserves the overall physical state. An illustration of a single step in the gauge fixing procedure is shown in \cref{fig:gauge_fixing}.

\section{Compressing MPS with \pbc}
\label{appendix:compression}
\begin{figure}[!htbp]
    \centering
    \includegraphics[width=\columnwidth]{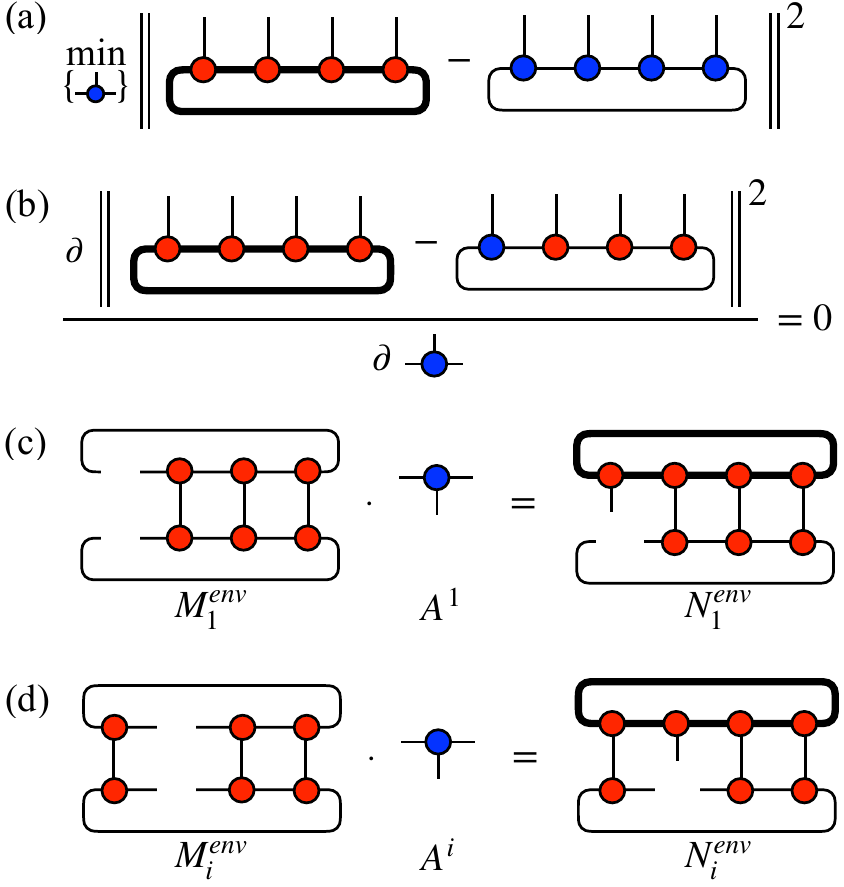}
    \caption{
     Compression process for MPS with \pbc, where the variational local tensors of the compressed MPS are colored in blue. (a) The compression is based on minimizing the squared Frobenius distance between the original MPS (symbolized with thick bonds) and the compressed MPS (symbolized with thin bonds) with respect to the local tensors of the compressed MPS. The optimization is performed by sequentially updating local tensors. (b) The first local tensor is determined by enforcing the variational extreme value conditions. (c) This leads to a linear equation for $A^{1}$, which can be solved by (pseudo-) inverting $M_1^{\text{env}}$. (d) The tensor update proceeds iteratively, sweeping the local tensors in the compressed MPS until convergence is achieved.
     }
     \label{fig:compressMPSpbc}
\end{figure}

This section explains the compression algorithm for MPS with \pbc to approximate it with an MPS of smaller bond dimension. The fundamental principle is shown in \cref{fig:compressMPSpbc}(a), where the uncompressed original MPS $|\psi_{0}\rangle$ is marked with thick bonds, while the compressed target MPS $|\psi(\{A^{i}\})\rangle$ is addressed with thin bonds for distinction. The compression is performed by minimizing the squared Frobenius distance $\mathrm{dist}$ between $|\psi_{0}\rangle$ and $|\psi(\{A^{i}\})\rangle$ 
\begin{equation}
    \mathrm{dist} = \langle \psi_{0} | \psi_{0} \rangle - 2\langle \psi_{0} | \psi(\{A^{i}\}) \rangle + \langle \psi(\{A^{i}\}) | \psi(\{A^{i}\}) \rangle.
\end{equation}
with respect to the local tensors $\{A^{i}\}$.  

We use a sequential update of the local tensors $\{A^{i}\}$ until convergence~\cite{verstraete2004renormalization}. The update equations for local tensors are derived based on the condition that the first-order partial derivative must be zero at extrema. To illustrate the process, we present the derivation and update procedure for the first local tensor $A^{1}$ in \cref{fig:compressMPSpbc}(b-c), where we consider a real-valued MPS and highlight the local tensor to be optimized in blue. From the condition that the first-order partial derivative is zero, as shown in \cref{fig:compressMPSpbc}(b), we derive the update equation for $A^{1}$ in \cref{fig:compressMPSpbc}(c) using diagrammatic language. To formalize this, we introduce the environments $M_{1}^{env} \equiv \frac{\partial^{2} \langle \Psi(\{A^{i}\}) | \Psi(\{A^{i}\}) \rangle}{2\partial (A^{1})^{2}}$ and $N_{1}^{env} \equiv \frac{\partial \langle \psi_{0} | \psi(\{A^{i}\}) \rangle}{\partial A^{1}}$ of $A^{1}$. These are equal to the contraction of the tensor network corresponding to the inner product $\langle \psi(\{A^{i}\}) | \psi(\{A^{i}\}) \rangle$ and $\langle \psi(\{A^{i}\}) | \Psi \rangle$, except for $A^{1}$. After permuting and reshaping $M_{1}^{env}$ and $N_{1}^{env}$ to matrix form, the update equation for $A^{1}$ reads as
\begin{equation}
    M_{1}^{env} \cdot A^{1} = N_{1}^{env}.
\end{equation}
Using the singular value decomposition
\begin{equation}
    M_{1}^{env} = U_{1} S_{1} V_{1}^{\dagger},
\end{equation}
the optimized tensor becomes
\begin{equation}
    A^{1} = V_{1} \bar{S_{1}}^{-1} U_{1}^{\dagger} N_{1}^{env},
\end{equation}
where $\bar{S_{1}}^{-1}$ is the pseudo-inverse of $S_{1}$. After updating $A_{1}$, the algorithm proceeds to update $A_{2}$ similarly. The optimization sweeps through all local tensors $\{A^{i}\}$ (see \cref{fig:compressMPSpbc}(d)), until $\mathrm{dist}$ remains unchanged within a predefined precision threshold.

\begin{figure}[!htbp]
    \centering
    \includegraphics[width=\columnwidth]{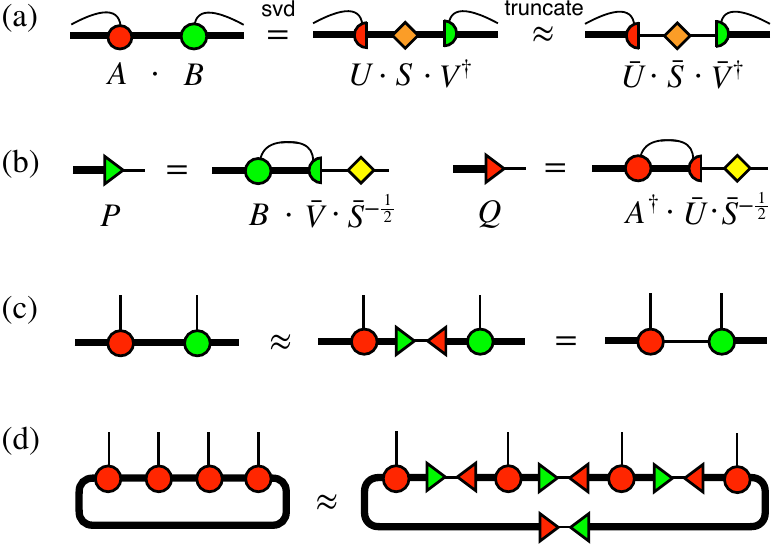}
    \caption{
     Processes for initializing local tensors of an MPS with smaller bond dimension by inserting pairs of isometry projectors into the original MPS. (a-b) Steps for determining a pair of isometry projectors $P$ and $Q$ for two neighboring local tensors $A$ and $B$ (colored red and green respectively). (a) After reshaping the tensors into matrices $A$ and $B$, a SVD decomposition $AB = USV^{\dagger} \approx \bar{U}\bar{S}\bar{V}^{\dagger}$ is performed, followed by truncation. (b) The projectors are determined as $P = B \bar{V}S^{-\frac{1}{2}}$ and $Q = A^{\dagger} \bar{U} \bar{S}^{-\frac{1}{2}}$. (c-d) Steps for approximating the original MPS with one having a smaller bond dimension. (c) Example of inserting the previously determined pair of projectors into the bond and contracting them with the local tensors. (d) Insertion of all pairs of projectors to obtain the MPS with the desired smaller bond dimension.
     }
     \label{fig:hosvd_projection}
\end{figure}

Before this optimization procedure, the compressed tensors are initialized using techniques from higher-order tensor renormalization group~\cite{xie2012coarse,yoshiyama2020higher}. For this, projectors are introduced between each pair of neighboring MPS tensors to compress the bond dimension based on a local optimization, as shown in \cref{fig:hosvd_projection}. In practice, this provides a good initial state for a stable optimization.

\section{Maximizing the success rate}
\label{appendix:max_success_rate}
Minimizing $C_{\alpha} C_{1-\alpha}$ directly maximizes the success rate $\mathrm{P}_{\mathrm{success}}$ of the probabilistic approach to incorporate \pbc, as given in \cref{eq:success_rate_final}. To identify the minimum of this product, we apply the Cauchy–Schwarz inequality
\begin{equation}
    \Bigg( \sum_{i=1}^{D} s^{2\alpha}_{i} \Bigg) \Bigg( \sum_{i=1}^{D} s^{2-2\alpha}_{i} \Bigg) \geq \Bigg( \sum_{i=1}^{D} s_{i} \Bigg)^{2}.
    \label{eq:CauchySchwarz}
\end{equation}
Equality is achieved if and only if the two vectors formed by $\{ s^{\alpha}_{i} \}$ and $\{ s^{1-\alpha}_{i} \}$ are linearly dependent. This condition is always satisfied when $\alpha = 1/2$. Hence, the product $C_{\alpha} C_{1-\alpha}$ is minimized at $\alpha = 1/2$, and the post-selection success probability $\mathrm{P}_{\mathrm{success}}$ is maximized in this case.

\section{Decomposition of bond matrix into local gates}
\label{appendix:bond_matrix_decomposition}
We explicitly construct a circuit that decomposes the qubit gate shown in \cref{fig:bondmatrix} into local gates. Let $D$ be the bond dimension of a given real-valued diagonal bond matrix, and assume $D = 2^{n}$ for some integer $n$.  Only the first column of $\Lambda^{\alpha}$ contributes to the output since it is applied to the initial state $|0\rangle^{\otimes n}$.\footnote{Similarly, only the first row of $\Lambda^{1-\alpha}$ is significant because of the post-selection process and the same arguments hold.} Therefore, it suffices to decompose the target unitary gate up to its first column. A key observation is that nonzero elements of $\Lambda^{\alpha}$ appear only when the state of the ancilla qubits matches that of the remaining qubits.
This pattern arises from the fact that the first column is a vector obtained by reshaping a diagonal matrix. Based on this structure, we first prepare the desired amplitudes, corresponding to the diagonal entries of the original matrix, on the first half of the qubits (ancillas). Then, we apply $n$ CNOT gates to ensure that nonzero elements appear only when the second half of the qubits matches the state of the ancilla qubits. This guarantees that the output amplitudes occupy the correct positions and thereby completes the preparation of the desired state.

To begin, we apply the Gram-Schmidt process to construct a unitary gate $U_{\mathrm{sub}}$ from the vector formed by the normalized diagonal elements, denoted $\{\tilde{s}_{i}\}$, which satisfy $\sum_{i=1}^{D} |\tilde{s}_{i}|^{2} = 1$. We then use Givens rotations~\cite{vartiainen2004efficient} to decompose $U_{\mathrm{sub}}$, retaining only the operations needed to generate its first column. Suppose
\begin{equation}
    U_{\mathrm{sub}} = 
    \begin{bmatrix}
        \tilde{s}^{(0)}_{1} & u^{(0)}_{1,2} & u^{(0)}_{1,3} & \cdots & u^{(0)}_{1,D} \\
        \tilde{s}^{(0)}_{2} & u^{(0)}_{2,2} & u^{(0)}_{2,3} & \cdots & u^{(0)}_{2,D} \\ 
        \vdots & \vdots & \vdots & \cdots & \vdots \\
        \tilde{s}^{(0)}_{D-1} & u^{(0)}_{D-1,2} & u^{(0)}_{D-1,3} & \cdots & u^{(0)}_{D-1,D} \\
        \tilde{s}^{(1)}_{D} & u^{(1)}_{D,2} & u^{(1)}_{D,3} & \cdots & u^{(1)}_{D,D}
    \end{bmatrix},
\end{equation}
where the upper index is added to $\{\tilde{s}_{i}\}$ for convenience in later discussion. Then, the goal of the Givens rotations $\{R_{k,1}\}$ is to eliminate all elements, one at a time, except for the one at the first index. According to the zero-out condition imposed by each rotation, we require that for each step $k$,
\begin{equation}
    R_{k,1} (\theta_{k,1}) \cdot
    \begin{bmatrix}
        \tilde{s}^{(0)}_{k-1} \\
        \tilde{s}^{(1)}_{k}
    \end{bmatrix}
    =
    \begin{bmatrix}
        \tilde{s}^{(1)}_{k-1} \\
        0
    \end{bmatrix},
    \label{eq:givens_rotation}
\end{equation}
where $R_{k,1} (\theta_{k,1})$ is a rotation-Y gate acting on the $(k-1)$-th and $k$-th components to set to zero the lower component. The equation determines the rotation angle 
\begin{equation}
    \theta_{k,1} = -2\mathrm{arctan}\Big( \frac{\tilde{s}^{(1)}_{k}}{\tilde{s}^{(0)}_{k-1}} \Big)
\end{equation}
and the updated value at step $k-1$ is given by
\begin{equation}
    \tilde{s}^{(1)}_{k-1} = \mathrm{cos}\bigg(\frac{\theta_{k,1}}{2}\bigg) \tilde{s}^{(0)}_{k-1} - \mathrm{sin}\bigg(\frac{\theta_{k,1}}{2}\bigg) \tilde{s}^{(1)}_{k}.
\end{equation}
These rotations are applied sequentially from the bottom up, in order to zero-out all elements below the first row. Importantly, Givens rotations preserve the inner product and thus the norm of the vector. As a result, the first element becomes $\pm 1$, since the original $\{\tilde{s}_{i}\}$ consists of real numbers and is normalized. If the first element becomes $-1$, it can be flipped to $+1$ by shifting $\theta_{2,1}$ to $\theta_{2,1}+2\pi$. Therefore, the decomposition can be achieved by the following equation
\begin{equation}
    \prod_{k=2}^{D} \big[ I_{k-2} \oplus R_{k,1}(\theta_{k,1}) \oplus I_{D-k} \big] \cdot U_{\mathrm{sub}} = I_{D},
\end{equation}
where $I_{k}$ denotes the $k \times k$ identity matrix. 

Next, we introduce the qubit-level realization of Givens rotations in the Gray code basis~\cite{vartiainen2004efficient}. When Gray code is used to map the matrix indices to qubit basis states, each Givens rotation corresponds to a multi-controlled rotation-Y gate, denoted as $C^{n}R_{Y}$, acting on qubits. Under the big-endian convention, the mapping from a matrix index $i \in \{1,2,\cdots,D\}$ to the corresponding Gray-coded qubit basis state involves two steps. First, the index $i$ is converted into its $n$-bit binary representation, resulting in the state $|b_{0} b_{1}\cdots b_{n-1}\rangle$, where $b_{0}$ is the most significant qubit. Second, the binary state is mapped to a Gray-coded state $\gamma_{n}(i-1) = |g_{0} g_{1} \cdots g_{n-1}\rangle$ according to the rule
\begin{equation}
    g_{i} = \left\{
            \begin{array}{ll}
                b_{0}, & \text{if } i = 0 \\
                b_{i-1} \oplus b_{i}, & \text{if } i > 0.
            \end{array}
            \right.
\end{equation}
This transformation ensures that any two adjacent indices involved in a Givens rotation correspond to basis states that differ by exactly one qubit. As a result, each Givens rotation can be implemented by a $C^{n}R_{Y}$ gate, where the target qubit is the one that differs between $\gamma_{n}(i-1)$ and $\gamma_{n}(i)$. The remaining qubits serve as control qubits conditioned on the positions where $\gamma_{n}(i-1)$ and $\gamma_{n}(i)$ agree. The rotation angle is determined by the Givens rotation. If the differing qubit states of the two indices involved in a Givens rotation appear in the order $|0\rangle$ and $|1\rangle$, the angle is applied directly. If the order is $|1\rangle$ and $|0\rangle$, the rotation angle must be taken with the opposite sign.. Once the full sequence of $C^{n}R_{Y}$ gates is determined, the decomposition of $U_{\mathrm{sub}}$ in the Gray code takes the form
\begin{equation}
     U_{\mathrm{sub}} = \prod_{k=2}^{D} \big[ I_{D-k} \oplus R_{D-k+2,1} (-\theta_{D-k+2,1}) \oplus I_{k-2} \big],
\end{equation}
where each term represents the Hermitian conjugate of a $C^{n}R_{Y}$ gate, and the product is ordered from left to right in increasing $k$. This construction requires $D-1$ $C^{n}R_{Y}$ gates in total. 

To complete the decomposition, we transform the Gray code basis back to the standard binary basis. This transformation is defined by the recursive relation
\begin{equation}
    b_{i} = \left\{
            \begin{array}{ll}
                g_{0}, & \text{if } i = 0 \\
                b_{i-1} \oplus g_{i}, & \text{if } i > 0
            \end{array}
            \right.
\end{equation}
and can be implemented using a sequence of CNOT gates. Specifically, for each $i=0$ to $n-2$, a CNOT gate is applied with $b_{i}$ as the controlled qubit and $b_{i+1}$ as the target. This step requires $n-1$ CNOT gates. Combined with the earlier Givens rotation decomposition in the Gray code basis, this transformation completes the construction of the qubit-level circuit representation for $U_{\mathrm{sub}}$. The entire circuit requires $D-1$ $C^{n}R_{Y}$ gates and $2n-1$ CNOT gates.

\bibliography{main}

\end{document}